\documentclass[%
 aps,
unsortedaddress,
 jmp,
 amsmath,amssymb,
 reprint,%
]{revtex4-2}

\preprint{APS/123-QED}

\usepackage[margin=1.5 cm]{geometry} 
\usepackage[]{algorithm2e}
\usepackage{graphicx}
\usepackage{dcolumn}
\usepackage{bm}
\usepackage[colorlinks,citecolor=blue,urlcolor=blue,bookmarks=false,hypertexnames=true]{hyperref}
\usepackage{comment}
\usepackage{svg}

\usepackage{algorithm2e}

\begin{document}

\preprint{AIP/123-QED}

\title[Clock synchronization with correlated photons]{Clock synchronization with correlated photons}
\author{Christopher Spiess}\email{christopher.spiess@iof.fraunhofer.de}
\affiliation{%
Friedrich Schiller University Jena, 
Fürstengraben 1, 07743 Jena
}%
\affiliation{ 
Fraunhofer Institute for Applied Optics and Precision Engineering, Albert-Einstein-Strasse 7, 07745 Jena
}%
\affiliation{Max Planck School of Photonics, Hans-Knöll-Straße 1, 07745 Jena 
}

\author{Sebastian Töpfer}%
\affiliation{ 
Fraunhofer Institute for Applied Optics and Precision Engineering, Albert-Einstein-Strasse 7, 07745 Jena
}%
\author{Sakshi Sharma}%
\affiliation{%
Friedrich Schiller University Jena, 
Fürstengraben 1, 07743 Jena
}%
\affiliation{ 
Fraunhofer Institute for Applied Optics and Precision Engineering, Albert-Einstein-Strasse 7, 07745 Jena
}%

\author{Andrej Kr\v{z}i\v{c}}%
\affiliation{%
Friedrich Schiller University Jena, 
Fürstengraben 1, 07743 Jena
}%
\affiliation{ 
Fraunhofer Institute for Applied Optics and Precision Engineering, Albert-Einstein-Strasse 7, 07745 Jena
}%
\affiliation{Max Planck School of Photonics, Hans Knöll Straße 1, 07745 Jena, Germany 
}

\author{Meritxell Cabrejo Ponce}%
\affiliation{%
Friedrich Schiller University Jena, 
Fürstengraben 1, 07743 Jena
}%
\affiliation{ 
Fraunhofer Institute for Applied Optics and Precision Engineering, Albert-Einstein-Strasse 7, 07745 Jena
}%
\affiliation{Max Planck School of Photonics, Hans Knöll Straße 1, 07745 Jena, Germany 
}

\author{Uday Chandrashekara}%
\affiliation{ 
Fraunhofer Institute for Applied Optics and Precision Engineering, Albert-Einstein-Strasse 7, 07745 Jena
}%

\author{Nico Lennart Döll}%
\affiliation{ 
Fraunhofer Institute for Applied Optics and Precision Engineering, Albert-Einstein-Strasse 7, 07745 Jena
}%

\author{Daniel Rieländer}%
\affiliation{ 
Fraunhofer Institute for Applied Optics and Precision Engineering, Albert-Einstein-Strasse 7, 07745 Jena
}%

\author{Fabian Steinlechner}\email{fabian.steinlechner@iof.fraunhofer.de}
\affiliation{ 
Fraunhofer Institute for Applied Optics and Precision Engineering, Albert-Einstein-Strasse 7, 07745 Jena
}%
\affiliation{%
Friedrich Schiller University Jena, 
Abbe Center of Photonics, Albert-Einstein-Strasse 6, 07745 Jena
}%

\date{\today}

\begin{abstract}
Event synchronisation is a ubiquitous task, with applications ranging from 5G technology to industrial automation and smart power grids. The emergence of quantum communication networks will further increase the demand for precise synchronization in the optical and electronic domains, which implies significant resource overhead, such as the requirement for ultrastable clocks or additional synchronization lasers. Here we show how temporal correlations of energy-time entangled photons may be harnessed for synchronisation in quantum networks. We achieve stable synchronisation jitter $<68\,$ps with as few as 44 correlated detection events per 100-ms data package and demonstrate feasibility in realistic emulated high-loss link scenarios, including atmospheric turbulence. In contrast to previous work, this is accomplished without any external timing reference and only simple crystal oscillators. Our approach replaces the optical and electronic transmission of timing signals with classical communication and computer-aided postprocessing. It can be easily integrated into a wide range of quantum communication networks and could pave the way to future applications in entanglement-based secure time transmission.
\end{abstract}

\maketitle

\newpage
\section{Introduction}
Secure transmission of information and reliable event synchronisation are key requirements in critical infrastructures \cite{Bellamy.1995,Narula.2018}, especially in power grids \cite{Phadke.1994}, financial networks \cite{Angel*.2014}, and cloud database services \cite{Corbett.2013}. Regarding information security, quantum key distribution (QKD) offers the unique proposition of encryption keys whose confidentiality can be lower bounded by the laws of physics \cite{Bennett.2014,Ekert.1991,Northup.2014,Gisin.2007}. Quantum communication is thus poised to become a backbone of a secure information infrastructure, with networks of many hundreds of fibre links established, and integration of ground-space relays over more than 1,000 kilometres already underway\cite{Chen.2021}.

Quantum communication networks, which involve tight timing budgets in the optical and electronic domains, are also a prime example for the need for accurate synchronisation of remote parties. In classical communication networks, the network time protocol (NTP) ensures synchronization with accuracy down to milliseconds \cite{Mills.2011}, and the Global Navigation Satellite System (GNSS) \cite{Berceau.2016} can reduce this to nanoseconds.  Modern high-bit-rate QKD systems, however, operate at picosecond timescales, and inevitably call for even better synchronization \cite{Tomita.2010}. In state-of-the-art QKD experiments , this is typically achieved by auxiliary pulsed lasers \cite{Yin.2020,Chen.2021,Liao.2017} or stable references \cite{Wengerowsky.2020}, like rubidium clocks \cite{Marcikic.2006,Shi.2020} or GNSS \cite{,Steinlechner.2017,Ursin.2007,Ecker.2021}. Specialized classical infrastructure together with the  White Rabbit protocol \cite{Dierikx.2016} may reach sufficient performance as well \cite{Wahl.2020}. All together, the requirement for auxiliary components create a  resource overhead that reduces the scalability \cite{Diamanti.2016} of networks in fiber and even applicability in space, due to the tight constraints on power consumption and weight \cite{Oi.2017,Kerstel.2018}. Moreover, much like the case in classical telecommunication networks, these requirements are likely to increase further when considering advanced network protocols, e.g., quantum teleportation \cite{Ren.2017,Valivarthi.2016} and entanglement swapping \cite{Tsujimoto.2018,Pan.1998}. 

Quantum clock synchronization protocols have been proposed to tackle synchronizing distant clocks \cite{Jozsa.2000,Chuang.2000,Giovannetti.2001_disp} in multi-partite network settings \cite{Krco.2002,Kong.2018} and with quantum enhancement beyond the possibilities of classical physics \cite{Giovannetti.2001}. To this end, investigations over the last years have focused on the exploitation of the temporal correlations of time-energy entangled photons \cite{Quan.2020,Quan.2016,Quan.2019,DAuria.2020}. When photon pairs originate from a common creation event, this gives rise to a narrow correlation peak in their arrival time at remote receivers. In other words, the location of this correlation peak indicates the average time delay between the receivers and can thus be used to extract the time and frequency differences between remote clocks \cite{Valencia.2004,Ho.2009}. The recent past has seen further advancement with experiments achieving exceptionally high precision with respect to a common frequency reference \cite{Yao.2012}: 600\,fs with correlated photons \cite{Quan.2020}, or down to 60\,fs using Hong–Ou–Mandel interference \cite{Quan.2016,Quan.2019,DAuria.2020}. In deployed long-distance links, synchronization jitters as low as a few tens of picoseconds have been demonstrated using ultraprecise rubidium clocks or GPS-disciplined clocks \cite{Ecker.2021,Steinlechner.2017,Lee.2019}. Without atomic frequency references, the timing jitter is orders of magnitude higher (e.g., the frequency offset for standard crystal oscillators may be up to 8 orders of magnitude higher \cite{TiradoAndres.2019,Bregni.1997}). This frequency difference results in a temporal drift of the correlation peak and makes it impossible to find the initial timing offset for synchronization. This has limited the applicability of synchronization based purely on quantum correlations \cite{Ho.2009}. Additional rubidium clocks, GPS-disciplined clocks, or other common time references have long been a necessary requirement for deploying long-distance and high-bit rate quantum communication systems \cite{Steinlechner.2017,Shi.2020,Ecker.2021,Wengerowsky.2020}. 

In this article, we build on the ground-breaking work of Ho et al. \cite{Ho.2009} and Valencia et al. \cite{Valencia.2004} and establish the feasibility of picosecond-level synchronization using the correlations of photon pairs in realistic link scenarios. Unlike state-of-the-art field experiments, which employ active electro-optic modulation to actively encode synchronization sequences \cite{CostantinoAgnesi.2020,Calderaro.2020,Williams.06.03.202112.03.2021}, we accomplish this purely in postprocessing, without any requirement for auxiliary hardware. Our protocol (section \ref{sec:synchronization_protocol}) estimates the clock frequency difference and compensates for the resulting broadening of correlation peaks (section \ref{sec:initialization}). This allows us to find the correlation peak, even in low signal-to-noise scenarios. Tracking the position of the correlation peak during the communication session allows us to correct for residual clock instabilities and achieve synchronization with root-mean-square (RMS) jitters $<68\,$ps  (section \ref{sec:live_tracking}). What is remarkable is that the approach also works when correlated detection events are as low as $440\pm200$\,cps, as is expected in real lossy links scenarios. Even more remarkably, these values are completely comparable to $30-50\,$ps jitters reported for systems that employ high-performance GPS-disciplined rubidium clocks \cite{Steinlechner.2017,Ecker.2021}. Finally, we establish the feasibility of the protocol for communication links through turbulent atmosphere, where channel fades have a major impact on the detected photons statistics (section \ref{sec:emulated_link}) \cite{GarciaLorenzo.2011}.

The results of our proof-of-concept experiment show that time-correlated photon pairs can be a valuable resource for synchronization with minimal hardware overhead. The methods work not only for laboratory experiments, but also for deployed QKD systems in high-loss link scenarios. In this approach, synchronization is a by-product of the key exchange without compromising the secure key rate. It provides a simple way of enhancing the timing resolution in distributed quantum information processing tasks. The core of the algorithm is not limited to correlated photon pairs and applies universally to any correlation features. In particular, it is readily extended to prepare-and-measure approaches \cite{Bennett.2014}. Our method for software-based synchronization in postprocessing can thus be implemented straightforwardly in state-of-the-art QKD systems and paves the way towards quantum networks \cite{Komar.2014} with improved synchronization performance, as well as entirely new applications such as secure time transfer \cite{Dai.2020,Lee.2019_attack}. 
\newline
\section{Results}

\subsection{Synchronization protocol}\label{sec:synchronization_protocol}
\begin{figure*}[htb]
	\centering
	\includegraphics[width=70.590mm]{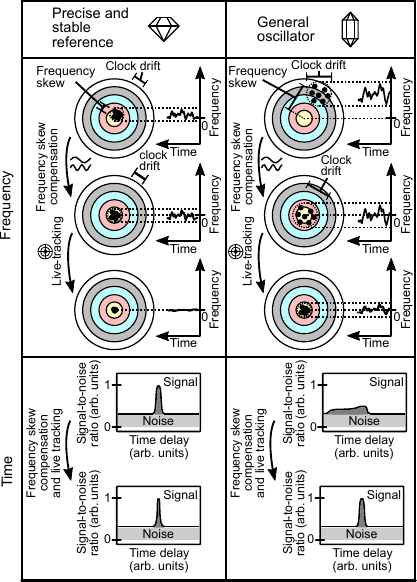}
	\caption[]{\textbf{Clock performance impacts.} The frequency of real, imperfect clocks differs from a nominal value and varies over time. The effect is much stronger for general oscillators (e.g. quartz oscillators) than precise and stable references (e.g. rubidium clocks). The compensation of frequency skew (i.e. frequency difference of clocks) increases the accuracy, and further enhancements to the precision are made through live tracking of the clock drift (i.e. time-varying clock frequencies). The correlation peak in time-domain shows significant timing jitter and may even be asymmetric, because of strong frequency variation over time in general oscillators. Clock frequency skew compensation and live tracking can reduce the additional timing jitter from poor-performing clocks to a minimum.}
	\label{fig:figure_clock_basics}
\end{figure*}

\begin{figure*}[htb]
	\centering
	\includegraphics[width=154.649mm]{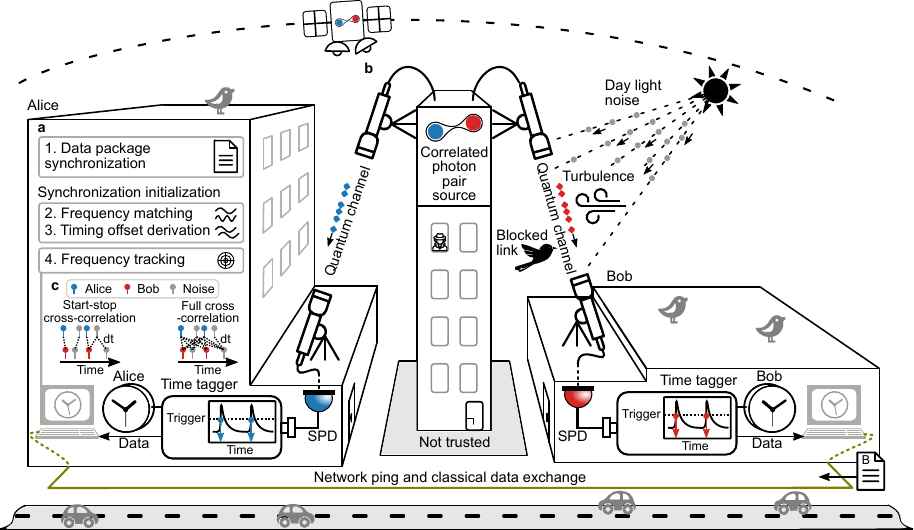}
	\caption[]{\textbf{Schematic synchronization workflow and derivation of timing information with correlated photons.} \textbf{a} Steps for synchronization include: data package synchronization, clock frequency skew estimation, timing offset calculation and correlation peak tracking over fast Fourier transform (FFT) or Start-Stop method-based cross-correlations. \textbf{b} Time-correlated photons induce electric pulses on single photon detectors (SPD) that are time-tagged when the voltage exceeds a threshold value. These correlated detection events act as mediator to synchronize the master and slave clock of two communicating parties Alice and Bob. The clocks may be the time tagger's internal quartz crystal oscillators, external rubidium or GPS-disciplined clocks. The synchronization performance is compared with a local reference from Bob for varying quantum channel attenuation and emulated turbulence with Fried parameter $r_0 = 1\,$mm and beam waist of 3.4\,mm (1/$e^2$) on the channel to Bob. \textbf{c} The cross-correlations are calculated through either FFTs or in a computation-efficient Start-Stop method. The former is crucial during initialization when the exact timing offset between Alice and Bob is unknown, i.e. a given timing offset uncertainty larger than the inverse single photon rate (see Methods \ref{SI:clock_skew_fine_tuning} for more information). Here, time differences (dt) between all time tags are included. In the Start-Stop method are only two subsequent time tags considered. }
	\label{fig:figure_exp_setup}
\end{figure*}

The performance of remote clocks used in communication networks has a great impact on the most appropriate method for their synchronization. High performance oscillators (e.g. Rubidium clocks) provide a precise frequency, but also exhibit small but constant difference frequencies (or clock frequency skew). In analogy to archery, the archer hits a spot on average, but off-center (see Fig. \ref{fig:figure_clock_basics}). These oscillators provide high accuracy, i.e., small clock drift, as well -- the frequency variation over time is small (the archer hits an aimed spot with small scatter). Both a small clock drift and clock frequency skew are necessary to maintain low synchronization timing jitters (the archer always hits the same spot that corresponds to the mark). Such precise and stable references relax the requirements on synchronization protocols and are still commonly employed in experiments \cite{Ecker.2021,Steinlechner.2017,Lee.2019}. The focus of the proposed work are general oscillators, which are widely used but exhibit a frequency skew that is orders of magnitude higher. Hence, any synchronization protocol for such oscillators will rely crucially on an initial compensation of frequency skew - especially in real link scenarios with low signal-to-noise ratio. Their strong and nonlinear drifts also make live tracking of the correlation with short feedback loop times necessary. Furthermore, the optical link characteristics play a crucial role, since it is the only mediator for timing information.
\newline

To this end, we have established a synchronization protocol, which consists of four main steps, where each step reduces the  timing uncertainty further - from coarse millisecond timing down to a picosecond level or smaller (Fig. \ref{fig:figure_exp_setup}a). In our experimental implementation, we use entangled photon pairs from an untrusted source as a mediator of timing signals between distant parties (Fig. \ref{fig:figure_exp_setup}b). The photons are produced at 810 nm via spontaneous parametric down-conversion with a 405-nm continuous-wave pump laser, and exhibit an intrinsic timing correlation jitter of $<2.5\,$ps (spectral bandwidth $0.4\pm0.17$\,nm). The first part of the correlation-based synchronization protocol is to align the data packages at the two receivers. The electronic time taggers generate time tags for single-photon detection events relative to their respective internal clocks. These time tags are stored on a personal computer in the form of data packages, each with an acquisition time of approximately 100\,ms. To detect the photons, we employ Si-SPAD detectors with a photon detection efficiency of approximately 60\,\%, dark count rates of 300\,cps and a timing jitter of approximately 140-180\,ps (RMS). As a very first step in the initialization, we use classical network pinging through the NTP \cite{Mills.2011}. This establishes coarse millisecond synchronization and ensures that the data packages to be compared carry correlated detection events - this procedure is not part of the paper. The next step is to identify the correlation peak in these data packages. However, due to the clock frequency skew, the position of the correlation peak is itself a function of time time, which results in significant spreading of the correlation peak over the 100\,ms integration time. The clock frequency skew could be computed by observing a moving correlation peak, which is a typical chicken-and-egg problem, as it implies that the peak has already been identified \cite{Ho.2009}. This becomes especially critical in low-signal scenarios, where it is impossible to recover correlation features (see Methods \ref{SI:ho_vs_our_work} for impact of signal and clock frequency skew on correlation peaks). To address this issue, we introduce a coarse clock frequency skew compensation before identifying the correlation peak with high visibility in the second step. This reduces the frequency skew of typical quartz oscillators from approx. 20\,$\mu$s/s to $\le$1\,$\mu$s/s, and thereby reduces the spreading of the correlation peak to $\le100\,$ns over the typical acquisition time of 100\,ms. By squeezing the correlation peak, it also enhances the signal-to-noise ratio, which makes synchronization in high-loss communication scenarios possible. In the third step, we cross-correlate the time tags of matched data packages using the fast Fourier transform (FFT) convolution and derive the precise time difference between the master and slave, i.e., the timing offset. Optionally, we fine-tune the clock frequency skew through calculating the correlation peak over the Start-Stop method \cite{Brunel.1999,Alleaume.2004,Martinez.2016}: a photon arriving at Alice (Bob) starts the counter and the second photon at Bob (Alice) stops it. The collection of time differences in a histogram represents the cross-correlation. This is computationally substantially more efficient than FFTs because it considers only two subsequent and neighboring time tags (as opposed to all, in the case of FFTs, see Fig. \ref{fig:figure_exp_setup}c). The downside of this cross-correlation method is that it requires a good initial estimate of the timing offset. Specifically, the precision of the timing offset must be smaller than the inverse single photon count rate (see Methods \ref{SI:clock_skew_fine_tuning} for more information). In summary, after data package synchronization (step 1), we match the clock frequencies in step 2 and estimate the timing offset in step 3. These are the requirements to start the quantum communication session. During a long communication session, the frequency of unstable clocks will change unpredictably. It follows a frequency mismatch between the two clocks that enormously increases the total system jitter. Live tracking of the correlation peak during the communication session and fast feedback loops in step 4 mitigates against fast variations of the clock frequencies. 
\newline

\subsection{Implementation of the synchronization protocol}\label{sec:implementation}
With the basic workflow established, we now consider the major challenges associated with its implementation in real link scenarios. Low signal-to-noise ratio and large clock frequency skews have a severe impact during synchronization initialization (section \ref{sec:initialization}). In worst-case scenarios, it is not possible to synchronize the system, as the correlation features stay hidden under noise. Initial sweeps of the clock frequency counteract this and enable successful initialization. Having taken this first hurdle, it is easy to keep the slave clock frequency locked to the master during the communication session. Here we take advantage of filtering the data in time by narrowing down the observation window of correlation features. In contrast to time windows as large as the acquisition time of 100\,ms during the initialization, it is narrowed down to a few tens of nanoseconds that reduces the noise drastically. However, clock drifts create a frequency mismatch between the master and the slave clock after the initialization of synchronization procedure. This increases the timing jitter during the communication session, which in turn increases the quantum bit error rate. All of this is circumvented by tracking of the clock frequency and immediate compensation as described in section \ref{sec:live_tracking}. 


\subsubsection{Initialization}\label{sec:initialization}

\begin{figure*}[htb]
	\centering
	\includegraphics[width=180.924mm]{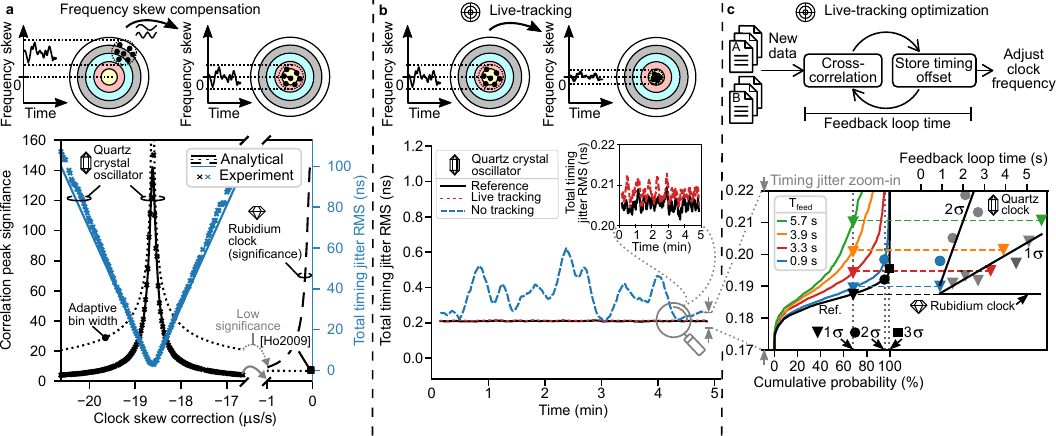}
	\caption[Second Figure]{\textbf{Clock frequency skew compensation and live tracking.} \textbf{a} The ratio of peak height to background standard deviation (significance) increases and the correlation peak spread reduces for correctly compensated clock frequency skew. Rubidium clocks do not require any compensation of the clock frequency skew and result in strong correlation peaks. The significance can be increased by adapting the bin width to the current clock frequency skew (see Methods \ref{SI:ho_vs_our_work}). The analytical trend lines (significance equation \ref{eq:trend_Sp_deltau} and timing jitter equation \ref{eq:trend_sigma_delta} in Methods \ref{SI:total_system_jitter}) are derived from the experimental maximum significance of 141. It provides the smallest residual clock frequency skew of 48\,ns/s and smallest timing jitter of 2.4\,ns over the acquisition time of 0.1\,seconds. \textbf{b} Live tracking of the correlation peak reduces the drift of clock frequency skew over time. The clock frequency skew is adjusted every 600\,ms in a feedback loop and result in much smaller total timing jitter that is almost identical to the reference with same clocks. The synchronization jitter contribution is as small as $35\pm8$\,ps with acquisition time of 100\,ms. Other experimental parameters in Methods \ref{SI:exp_photon_pair_rates}. \textbf{c} The timing offset from the cross-correlation peak location is stored for every incoming data package up to the feedback loop time. The instantaneous clock frequency difference is equal to dividing the average timing offset by the passed time. Whereas the feedback loop time does not affect much the performance of stable rubidium oscillators, it heavily affects the synchronization jitters achieved with quartz oscillators. Here we characterize the performance by accumulating the occurrence probability for all jitter values during a 5-minutes session. The 1$\sigma$ (2$\sigma$, 3$\sigma$) timing jitter represents the maximum jitter in 68.3\,\% (95.5\,\%, 99.7\,\%) of time during the session. The synchronization jitter increases linearly with the feedback loop time (equation \ref{eq:sigma_drift}). Grey data points correspond to measurements that are not displayed in the cumulative probability plot.}
	\label{fig:figure_2}
\end{figure*} 
Accumulation of simultaneous photon detection events from matched data packages results in a correlation peak that is located at the timing offset between two communicating parties. Distinguishing the peak correlation that corresponds to this offset from the noisy background is a major challenge. For that, we introduce the statistical significance as the peak height normalized to the noisy background standard deviation \cite{Ho.2009}. Strong spread of the correlation peak in time reduces the significance. This spread is equivalent and characterized by the total system timing jitter, including detection, time tagging, and synchronization jitter. Typical root-mean-squared (RMS) timing jitters of single photon detection systems are smaller than 500\,ps and also time tagging units achieve jitters smaller than a few tens of picoseconds with ease.  Poor synchronization on the other hand, with its corresponding RMS synchronization jitter $\sigma_{\text{sync}}$, has the strongest impact on the total system jitter. It may amount to 1\,$\mu$s, with uncorrected clock frequency skew of 20\,$\mu$s/s and an acquisition time of 100\,ms. The synchronization jitter is time-dependent and increases with acquisition time $T_a$ and uncorrected clock frequency skew $\Delta u$,

\begin{equation}
    \sigma_{\text{sync}} = \frac{1}{2} \Delta u T_a. \label{eq:sync_jitter}
\end{equation}
\noindent
Frequency skew reduces the height of the correlation peak and thereby decreases the probability to correctly identify it. Systems with stable oscillators, like rubidium clocks, do not suffer from a reduction of significance (Fig. \ref{fig:figure_2}a) without correction. General crystal oscillators, on the other hand, lower the signal close to the noise level and thus make it impossible to find the correlation peak. To counteract this, we introduce the compensation of skew, so that the $i^\text{th}$ time tag on Bob's side $t_i$ is corrected to $t_i^\text{c}$,
\begin{equation}
t_i^c = t_i - \Delta u_c\left(t_i-t_{0}\right). \label{eq:skew_correction}
\end{equation}
\noindent
For different values of corrections $\Delta u_\text{c}$ we calculate the cross-correlation under the constraint of computation effort, limiting us to the computation of approximately 280 cross-correlation functions (see Methods \ref{SI:cross_correlation_computation_and_algorithm} for the algorithm and computational requirements). This is sufficient to search in a clock frequency skew range from -20 to $+20\,\mu$s/s (decided according to our quartz crystal clock). The step size of $0.14\,\mu$s/s provides clock frequency skews with an uncertainty $<0.14\,\mu$s/s. Optimum clock frequency skew is indicated by the maximum significance of the correlation peak in the search window (Fig. \ref{fig:figure_2}a). The maximum significance of 142 allows to create helpful trend lines with the coincidence rate 7\,kcps, single rate on Alice's side 244\,kcps, Bob's side 232\,kcps and acquisition time of 0.1\,seconds (see Methods \ref{SI:total_system_jitter}). The experimental timing jitter is derived by fitting the correlation peak with a flat-top Gaussian function. Note that the significance enhances by adapting the bin size to the correlation peak spread (see Methods \ref{SI:ho_vs_our_work}) and reaches values of up to 6.8 (see Methods \ref{SI:ho_vs_our_work}, equation \ref{eq:ho_S_p}) without clock frequency skew compensation. However, the correlation peak spread is unknown at this point of time, as the true clock frequency is uncertain. Such a random selection of the bin size reduces the significance again. In contrast to previous work \cite{Ho.2009}, we could increase the significance by a factor of $\sqrt{\Delta u / \delta u} \approx 20$, thanks to our new method of clock frequency skew compensation. Further reduction of the residual clock frequency skew is made by a more computation-efficient fine-tuning (see Methods \ref{SI:clock_skew_fine_tuning}) or by observing moving correlation features \cite{Ho.2009}. Whereas we cannot avoid noise, clock frequency skew compensation sharpens the correlation features and creates higher visibility of them (more information on the signal-to-noise ratio dependence in Methods \ref{SI:ho_vs_our_work}, Fig. \ref{fig:SI_significance_vs_SNR}). This provides noise resistance and makes it feasible to initialize low signal-to-noise ratio links.  
\newline

\subsubsection{Live tracking} \label{sec:live_tracking}

\begin{table*}[htb]
\caption{\label{tab:table1}Residual synchronization jitter root-mean-square (RMS) for different link rates and coincidence to-accidentals ratios (CAR) during live tracking. Considering the synchronization initialization, see Methods \ref{SI:ho_vs_our_work}, Fig. \ref{fig:HoSteinEcker_vs_us} for comparison of the different setups. The clocks are external crystal oscillators (XO) or clocks locked by the Global Navigation Satellite System (GNSS). \textbf{a} Low signal experiment, including satellite up-link turbulence (see Methods \ref{SI:high_loss_live_tracking}), \textbf{b} Moderate signal experiment, including satellite up-link turbulence (Fig. \ref{fig:figure_3}), \textbf{c} 1.2-km free-space and high-dimensional link experiment across Vienna by Steinlechner et. al. \cite{Steinlechner.2017} - synchronization jitter estimated from timing offset variation, \textbf{d} 143-km free-space link experiment between the Canary Islands La Palma and Tenerife by Ecker et. al \cite{Ecker.2021} - synchronization jitter estimated from timing offset variation. The coincidence window (half-width) depend on the detection system and is \textbf{a-b} 0.27\,ns \textbf{c} 1\,ns, \textbf{d} 0.5\,ns. *Only live tracking - synchronization initialization at stronger signal  }
\begin{tabular}{ |p{4.5cm}|p{2cm}|p{3cm}|p{3cm}|p{3cm}|  }
\hline
Parameter& \textbf{a} Low signal  &\textbf{b} Moderate signal  & \textbf{c} 1.2-km link in Vienna & \textbf{d} 143-km link between Canary islands\\
\hline
Clock type & External XO & External XO &GNSS&GNSS\\
Count rate Alice (kcps) &  $165 \pm 3$& $195 \pm 3$ &400&13 300 \\
Count rate Bob (kcps) &  $437 \pm 6$& $15 \pm 5$ &100&10\\
Correlation event rate (cps) &  $430 \pm 160$ & $440 \pm 200$&20 000&$>300$\\
CAR &  $10\pm4$& $84\pm13$&250&$<5$\\
Synchronization jitter RMS (ps) &  $98\pm6^*$& $68\pm8$ &$\approx 33 $&$\approx 50$\\
Clock drift RMS (ps/s$^2$) &  320& 320&$<5\,$ (typical)&$<5\,$ (typical)\\
Clock frequency skew (ppm) &  19& 19 &$10^{-6}$ (typical)&$10^{-6}$ (typical)\\
\hline
\end{tabular}
\end{table*}

At this point in the protocol, the sender and receiver have matched clock frequencies and the corresponding timing offset. The next challenge is to maintain this common time basis over extended periods via live tracking of the correlation peak. In the case of rubidium oscillators or GPS-disciplined clocks, which do not exhibit significant clock frequency drifts, this step is not necessary. Typical quartz oscillators, however, require the initial estimate for the clock frequency skew to be adapted in real time. In other words, we adjust the time-varying slave clock frequency to match the changed master clock frequency. These drifts result in a measurable change of the position of the correlation peak. The location is continuously tracked, and the slave clock frequency is afterwards adapted by the peak displacement over the elapsed time (see detailed algorithm in Methods \ref{SI:live_tracking_algorith}). In our experiment, the acquisition time is 100\,ms and feedback loop time for adapting the instantaneous clock frequency is 600\,ms. Figure \ref{fig:figure_2}b compares the total timing jitter with and without live tracking to Bob's reference that indicates zero synchronization jitter (see Methods \ref{SI:exp_photon_pair_rates} for the single count rate). Over a time window of 5\,min we find maximum clock frequency skew variations of up to $\pm 11\,$ns/s that show the importance of live-correcting these clock drifts of 320\,ps/s$^2$ (see Methods \ref{SI:experimental_clock_drift}). Without live tracking, the timing jitter would increase from 200\,ps to almost 700\,ps with 100\,ms acquisition time. Stable frequency references (here: rubidium clocks) introduce very low timing jitters of smaller than 1\,ps as consequence of live tracking (see Methods \ref{SI:live_tracking_algorith}). Quartz oscillators require more rigorous treatment and optimization of the algorithms. With a stronger drift of the clock $\partial (\Delta u) / \partial t$, there is timing jitter $ \sigma_\text{drift}$ introduced with longer feedback loop times $T_\text{feed}$ and acquisition time $T_a$, 

\begin{equation}
    \sigma_\text{drift} =  \dfrac{1}{2} \dfrac{\partial (\Delta u)}{\partial t} T_\text{feed}T_a.  \label{eq:sigma_drift}
\end{equation}
\noindent
Large feedback loop times compensate the fast changing clock frequencies not early enough and results in a linear increase of the overall timing jitter (Fig. \ref{fig:figure_2}c). Low signal and other link conditions reduce the synchronization quality further. This is part of the following section \ref{sec:emulated_link}. 
\newline

\subsection{Experiment under emulated link conditions}\label{sec:emulated_link}

\begin{figure*}[htb]
	\centering
    \includegraphics[width=178.232mm]{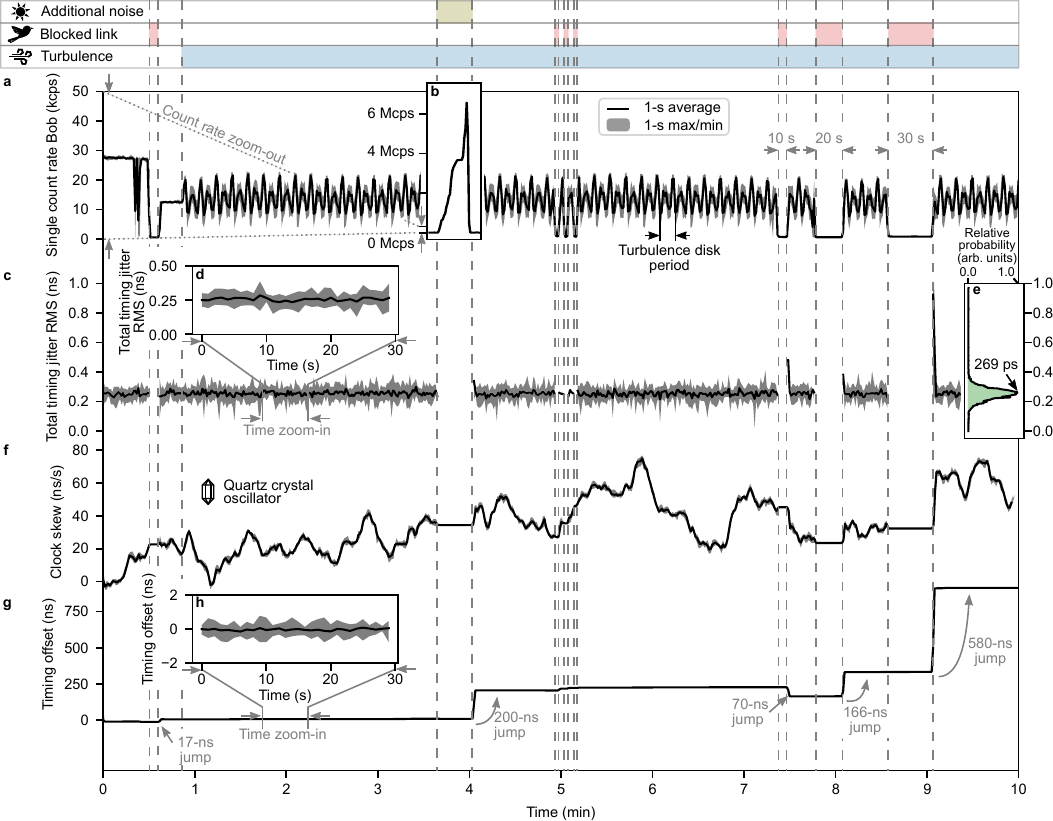}
	\caption[Second Figure]{\textbf{In-lab emulated free-space link experiment.} At different times we add additional noise at Bob's receiver, block the optical link or add turbulence (Fried parameter 1\,mm, $1/e^2$ full-width beam waist of 3.4\,mm). \textbf{a} Single count rate at the receiver Bob. The mean single count rate is $15\pm5$\,kcps with coincidence rate of $440\pm200\,$cps (coincidence window is equal to RMS timing jitter of 269\,ps) in the time window from minute 1 to 3.5. The count rate at Alice lies constantly at $195\pm3\,$kcps. \textbf{b} Introduction of a noise source increases Bob's count rate to 6\,Mcps. Low signal-to-noise ratio stops the correlation peak tracking and keeps the instantaneous clock frequency skew for correction constant. \textbf{c} The time-dependent total timing jitter root-mean-squared (RMS) remains constant (\textbf{d}) even after switching on the turbulence at minute 1. \textbf{e} The timing jitter probability distribution describes a Gaussian with its fitted center at $268.7\pm0.7$\,ps (minute 1 to 3.5). Setups with the same clock (perfect synchronization) show timing jitters of $260\pm2$\,ps from a fit to 10\,s of data acquisition, estimating the synchronization jitter RMS to $68\pm8$\,ps. \textbf{f} The tracking algorithm is stopped in regions with low signal-to-noise ratios (blocked optical link or additional noise) and results in constant clock frequency skew for correction there. \textbf{g} The timing offset jumps after low signal-to-noise ratios time intervals, as the correlation peak is not tracked and the clock frequency not adapted. This phenomenon is stronger for longer time of blocked links. \textbf{h} The timing offset varies within a window of $\pm1\,$ns on time scales of 30\,s. The feedback loop time and data package size amounts to 200\,ms and 100\,ms, respectively. Summary of this figure in table \ref{tab:table1}.}
	\label{fig:figure_3}
\end{figure*}

To confirm the feasibility of our approach in real-world link conditions, we perform a series of experiments with added background noise, and signal fades, as is expected in free-space links with atmospheric turbulence. To this end, we emulate several link scenarios and introduce turbulence via a rotating phase plate, designed to represent an atmospheric uplink to a satellite from the Mt. Teide optical ground station (Fried parameter is 1\,mm with $1/e^2$ full-width beam waist of 3.4\,mm  \cite{GarciaLorenzo.2011}). This turbulence results in strong variations of the  detected count rate of Bob (Fig. \ref{fig:figure_3}a). The rate at Alice, who is assumed to be co-located with the quantum source, remains constant. To test the algorithm's stability under elevated noise levels (for example, stray light), we introduced background counts via a light-emitting diode (see Fig. \ref{fig:figure_3}b). Even under these unfavourable conditions, live tracking with short feedback loops (200\,ms) ensures synchronization RMS timing jitters of $68\pm8$. This is comparable to 50\,ps synchronization jitter from systems with GPS-disciplined clocks \cite{Ecker.2021}. Our timing jitter is estimated with the same-clock timing jitter RMS of $260\pm2$\,ps and the average timing jitter of $268.7\pm0.7$\,ps after a Gauss fit to the jitter values from minute 1-3.5 (Fig. \ref{fig:figure_3}c-e). The correlation peak tracking algorithm stops when reaching a signal-to-noise ratio threshold that is set to 5. Setting any threshold is required, as it avoids that noise wrongly identifies as a correlation feature. Both the increased noise and interruptions of the optical link (e.g., due to objects crossing the beam path) stop the tracking, as the signal-to-noise threshold is not reached and leave the instantaneous clock frequency skew constant (Fig. \ref{fig:figure_3}f). This impacts only very lossy links with an increased synchronization timing jitter RMS to 98\,ps, as described in Methods \ref{SI:high_loss_live_tracking}. When the clock frequency skew for correction is not adapted in low signal regions, the correlation peak will start to move in time. As a consequence, the timing offset jumps to the value of the current location, as soon as the signal is back. Long periods of broken links follow in large timing offset jumps and depend on the clock's stability. These timing offset jumps are unique to unstable clocks. They require sufficiently large observation windows to not lose the correlation peak that is essential for synchronization. The timing offset is bound to fluctuations in a window of $\pm1\,$ns (Fig. \ref{fig:figure_3}g-h) in conditions of good signal (coincidence-to-accidentals ratio $>5$). In the Methods \ref{SI:high_loss_live_tracking} we present stable live tracking for much lower coincidence-to-accidentals ratios of only 10 (Table \ref{tab:table1}), comparable to high-loss link scenarios \cite{Ecker.2021}. The synchronization jitter RMS amounts to $98\pm6\,$ps with only 43 correlation events in a 100\,ms data package and 200\,ms feedback loop time. Note that the correlation peak significance during synchronization initialization is extremely low with the rates given, similar to \cite{Ecker.2021} (see Methods \ref{SI:ho_vs_our_work}, Fig. \ref{fig:HoSteinEcker_vs_us} for comparison of different setups in literature). Our synchronization initialization algorithms fail here, due to limited computation time and power that could not compensate the clock frequency skew enough for sufficient significant correlation features. Therefore, we shortly increase the signal-to-noise ratio during initialization and reduce it again later. In conclusion, the combination of a) adaptive correlation observation window, b) introduction of a coincidence-to-accidental ratio threshold, and c) short feedback loop times enable very stable operation and only a minor increase of the total timing jitters.

\section{Discussion}
These results clearly show that the compensation of clock frequency skew and live tracking of the frequency are suitable for application in a variety of real-world link scenarios. We could show that very few correlation events are already sufficient to establish and keep a synchronized quantum communication session with timing jitter in the range of tens of picoseconds over time scales of 10 minutes. Recently, we also confirmed these results on a 1.7\,km intra-city link. Reference \cite{Ho.2009} predicts that sufficient significance of larger than $\,7$ can only be achieved by clock frequency skews of $<1.3\,\mu$s/s in our realistic link experiment (Table \ref{tab:table1}b, Equation \ref{eq:ho_S_p} in Methods \ref{SI:total_system_jitter}). This means that the synchronization initialization would fail as our clock frequency skew amounts to $18.5\,\mu s$/s (see Methods \ref{SI:ho_vs_our_work}, Fig. \ref{fig:HoSteinEcker_vs_us} for comparison of different setups from other publications). It is a consequence of the poor visibility of the correlation peak. Furthermore, previous treatments have been limited to high signal-to-noise ratios. Here we demonstrate successful synchronization for much lower signal-to-noise ratios thanks to compensation with an effective frequency skew of smaller than 48\,ns/s that enables the enhancement of the significance by a factor of 20 as part of the initialization (Fig. \ref{fig:figure_2}a). This brings correlated photons and quartz oscillators in a much better position - not only for today's typical quantum communication schemes \cite{Steinlechner.2017}, but also for tomorrow's long-distance and low-signal scenarios. The intrinsic timing relations of correlated photons clearly have the potential to replace bulky external high-precision synchronization schemes and move resources from hardware to software. 
\newline

Finding the timing offset is crucial to start communication, but the low signal-to-noise ratio will reduce the correlation peak significance. Especially high-loss scenarios of $>46.9\,$dB \cite{Ecker.2021} (Table \ref{tab:table1}d) call for very low clock frequency differences for sufficient correction peak visibility. This typically requires highly precise oscillators, like rubidium or GPS-disciplined clocks. Quartz clocks are acceptable in principle, but demand a precise slave clock frequency sweep to achieve a frequency match with the master. Within our presented search window $-20\,\mu$s/s to $+20\,\mu$s/s it is impossible to find this precise frequency to the required accuracy of $<14\,$ns/s. High computation times give rise to limitations (see Methods \ref{SI:cross_correlation_computation_and_algorithm}). The solution is the prior knowledge of the approximate clock frequencies to narrow down the search window - ideally there will be just a single initial cross-correlation necessary to find the correlation peak. We propose to minimize the search window by a) more accurate frequency specifications from the clock manufacturer, b) calibration of clocks with high signal rates, or c) finding the clock frequency at a time, when computation limits do not constrain. High-performance computers with high parallel processing speeds, for example, with a graphical processing unit, reduce the computation time as well. Note that finding the initial clock frequency skew is a single-time issue, as it will not occur during high-duty quantum communication sessions. The algorithm tracks the aging of the clocks and adjusts the clock frequency continuously. Starting with large frequency uncertainty, the estimations improve over time. This provides a narrow clock frequency search range before every session, where the computation effort reduces drastically down to a few seconds – even for lossy long distant links. 
\newline

Systems with very high losses or accelerating clocks suffer from increased synchronization jitters during communication. Low number of correlation events increase the uncertainty of the clock frequency skew for compensation and unlock live-tracking feedback loops. Low clock stability further increases the synchronization timing jitter. High precision clocks clearly have an advantage here that allow for much lower synchronization jitters $<1\,$ps during communication sessions, due to low drifts through their high stability. Apart from signal-to-noise ratio and clock stability, the detector jitter is another factor for high synchronization performance. Particularly, high loss scenarios benefit from low detection jitters that increase the signal-to-noise ratio. We demonstrated that even 44 correlation events in 100\,ms data packages are sufficient for a synchronization jitter of a few tens of picoseconds. The event rate and signal-to-noise ratios are comparable to extremely high loss scenarios \cite{Ecker.2021} (Table \ref{tab:table1}d), but are enough to enable correlation peak tracking and keep the system locked. As nanowire detectors already have timing jitters of as low as tens of picoseconds, the synchronization performance with correlated photons will only increase in future.
\newline

To set our approach into a broader perspective, we also shortly discuss limits to moving objects, like satellite- \cite{Yin.2020,Liao.2017} or emerging drone-based \cite{SamanthaIsaac.2020,Conrad.06.03.202112.03.2021} quantum communication. Quantum sources in space provide a great platform to test and measure space-time effects on quantum communication protocols \cite{Bruschi.2014,Bruschi.2021} or perform high precision metrology \cite{Bruschi.20142,Kohlrus.2019}. Satellites introduce an effective clock frequency skew, due to the Doppler Effect caused by varying distance to the observer. The normalized Doppler shift $\overrightarrow{v}(t)/c$ for low earth orbit satellites increases from 0 to $2\times10^{-5}$ over a time scale of 6\,min, where $\overrightarrow{v}(t)$ is the satellite's time-dependent relative velocity and $c$ is the speed of light \cite{Ali.1998}. Equivalently, this creates a clock frequency skew that varies from 0 to a maximum of $\Delta u = 20\,\mu$s/s. Whereas the maximum clock frequency skew is comparable to our crystal oscillator, the clock acceleration $\partial (\Delta u)/\partial t$ is orders of magnitude bigger and amounts to approximately 55\,ns/s$^2$ (20\,$\mu$s/s divided by 6\,min). Thus, it is desirable to reduce the feedback and acquisition time to 100\,ms or smaller that could result in clock drift jitters of 265\,ps (equation \ref{eq:sigma_drift}). However, with only a few correlation events per second available \cite{Yin.2020,Yin1140}, it is impossible to choose acquisition times of $<1$\,second and still have sufficient signal-to-noise ratio. Using only correlated photons, synchronization is not possible without knowledge of the satellite's orbit. Sources with kcps rates \cite{Dai.2020}, on the other hand, give the opportunity to select small feedback cycles of 100\,ms and still have correlation events comparable to our experiment (Table \ref{tab:table1}b) that could be used to synchronize clocks with our method. Drones move much slower than satellites and may also be suitable for synchronization -- at least classically \cite{Bergeron.2019}. The speed is up to 30\,m/s, introducing clock frequency skews of up to 100\,ns/s (30\,m/s divided by the speed of light), being much smaller than the clock frequency skew of our crystal oscillators. The main concern is acceleration of the drone, reaching up to 7\,g (gravitational constant) and translating to 228\,ns/s$^2$ clock drift. Sufficient kcps coincidence rates \cite{SamanthaIsaac.2020} give chance to select short feedback cycles for the compensation of high drifts during drone acceleration and make synchronization feasible. With small feedback loops, correlation events open doors for live remote detection of velocity and acceleration of moving objects. 
\newline

In conclusion, correlated photons are great timing carriers, come to quantum communication systems naturally, and are easy to recycle for high performance synchronization down to a few tens of picoseconds. Today's point-to-point or lab-to-lab communication sessions will be integrated in a network with multiple users tomorrow, as can be found in action already \cite{Chen.2021}. High scalability, integration and fewer resources characterize the networks, where highly stable but bulky clocks should be an exception. Where correlated photons used to be inappropriate for synchronizing clocks with high skew and strong drifts \cite{Ho.2009}, we showed stable operation by new synchronization methods. Clock frequency skew compensation and correlation peak live tracking allows for a wider range of cases - especially in terms of scalability. Resistance to high losses, as would be common in large networks, still enables synchronization RMS jitters of $<68\,$ps and presents feasibility for application in real-life communication scenarios. Single photons are not copyable and bit-errors during communication are easy to detect due to the quantum origin of the single photon detection events \cite{Bennett.2014,Ekert.1991}, indicating furthermore the potential for quantum secured time transfer \cite{Dai.2020,Troupe.27.01.201801.02.2018,Lee.2019_attack}.

\begin{acknowledgments}
This research was conducted within the scope of the project QuNET, funded by the German Federal Ministry of Education and Research (BMBF) in the context of the federal government’s research framework in IT-security “Digital. Secure. Sovereign.”. Christopher Spiess is part of the Max Planck School of Photonics supported by BMBF, Max Planck Society, and Fraunhofer Society.
\end{acknowledgments}

\section*{Author Contributions}
C.S. designed the experiments with guidance from D.R.. C.S. performed the experiments. S.T. and S.S. developed essential hardware and software components with support and guidance from D.R.: S.S. and U.C. developed the entanglement source. C.S. wrote the main part of the Python processing script with assistance from S.T.. A.K. developed the turbulence testing setup with assistance from N.L.D.. F.S. proposed and directed the research. The first draft of the manuscript was written by C.S., F.S. and D.R. with assistance by M.C.P.. All authors discussed the results and reviewed the manuscript.

\section*{Competing Interests statement}
The authors declare no competing interest.

\newpage
\section{Methods}

\subsection{Correlation peak significance and comparison to previous work}\label{SI:ho_vs_our_work}

The peak significance $S_p$ of correlation features reduces with high clock frequency differences $\Delta u$, as described in \cite{Ho.2009},

\begin{equation}
    S_p = \sqrt{\dfrac{r_C^2 N}{r_A r_B}} =  \sqrt{\dfrac{r_C^2}{r_A r_B \Delta u}}, \label{eq:ho_S_p}
\end{equation}

\noindent
with coincidence (or signal) count rates $r_C$ and single rates from Alice $r_A$ and Bob $r_B$. This, however, only considers the situation that the bin width $\delta t = T_a / N $, with the number of bins $N$, has been perfectly adapted to the spread of the correlation peak $\delta t_{\text{spread}}$ over the acquisition time $T_a$,

\begin{equation}
    \delta t_{\text{spread}} = T_a \Delta u.
\end{equation}

\begin{figure}[htb]
	\centering
	\includegraphics[width=67.156mm]{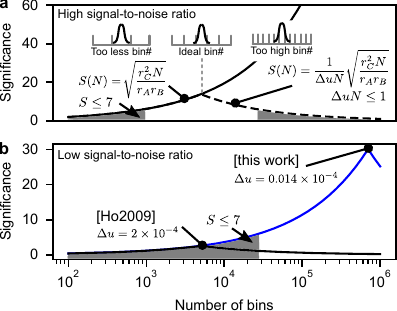}
	\caption{\textbf{Comparison of Ho2009 \cite{Ho.2009} and this work.} \textbf{a} Time tags are recorded over the acquisition time and sorted in a number of bins (bin\#) that impacts the significance. With single count rates $r_A = r_B \approx 77\,$kcps, signal rates $r_C \approx 15\,$kcps and clock frequency skew $\Delta u = 2\times10^{-4}$, the significance reaches 13 - sufficient for synchronization initialisation. \textbf{b} Methods by \cite{Ho.2009} fail for lower signal-to noise ratios, $r_A = 200\,$kcps, $r_B = 800\,$kcps, $r_C = 14\,$kcps, as the the correlation peak can not be found due to low significance of 2.5. Here we introduce compensation of the clock frequency skew that provides a better guess of the clock frequency skew, following in higher significance by the square root of number of cross-correlations in the clock frequency skew search (here is a factor of 12 improvement).}
	\label{fig:SI_Ho_vs_this_work}
\end{figure}

\noindent
The clock frequency skew $\Delta u$ is not know usually, so that it will be hard to guess the bin size correctly. This means reduction  of significance by $\sqrt{N}$ for too large bin widths (equation \ref{eq:ho_S_p}). Similarly, if the bin width was too small, signal would be distributed over several bins. The number of coincidences per bin will reduce by the ratio $\delta t_{\text{spread}} / \delta t$, 

\begin{equation}
    S(N) = \dfrac{1}{\Delta u N} \sqrt{\dfrac{r_C^2 N}{r_A r_B}},
\end{equation}

\noindent
with $\Delta u N \geq 1$. Figure \ref{fig:SI_Ho_vs_this_work}a depicts this behavior for the photon pair source from \cite{Ho.2009}. With their high signal-to-noise ratio, it is possible to find the very first correlation peak and determine the clock frequency skew from the correlation peak displacement over time. In this paper, we have tested lower signal-to-noise ratios that would result in a significance of barely 2.5 with a clock frequency skew of $\Delta u = 2\times 10^{-4}$ (Fig. \ref{fig:SI_Ho_vs_this_work}b). As a consequence, it would not even be possible to start with the algorithms as described in \cite{Ho.2009}. In this work, we propose a clock frequency skew compensation for this crucial initial step in low signal-to-noise environments. We find that improvements of the clock frequency skew uncertainty by a factor of 140 are feasible, depending on the fast Fourier transform - run times (see Methods \ref{SI:cross_correlation_computation_and_algorithm} for computational requirements). This provides an improvement of the significance by a factor $\sqrt{140} \approx 12 $ to a value close to 30 and enables reliable identification of the correlation peak even under low signal-to-noise ratios. Note that we reduce the residual clock frequency skew to $\delta u = 48\,$ns/s (with the rate given and its final peak significance) that even provide a significance improvement by a factor of 20.

\begin{figure*}[htb]
	\centering
	\includegraphics[width=107.715mm]{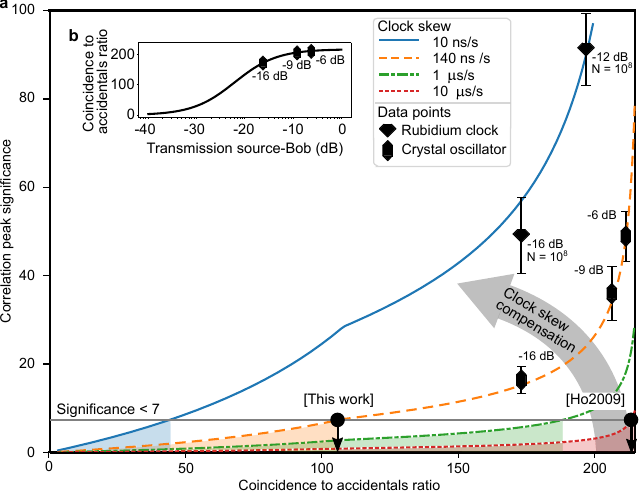}
	\caption{\textbf{Correlation peak significance for different loss scenarios.} \textbf{a} Estimation of correlation peak significance under different coincidence-to-accidentals ratios (loss) and clock frequency skews $\Delta u$. The experimental data points base on initial lossless rates of $R_A = 200\,\text{kcps}, R_B = 800\,\text{kcps}, R_C = 14\,\text{kcps}, r_\text{back} = 5\,\text{kcps}$ and maximum number of bins in the cross-correlation $N = 10^8$. The number of bins are adapted to the correlation peak spread, caused by the residual clock frequency skew, with $\Delta u = 1/N$. Error bars indicate standard deviation after slight variation of bin sizes by $\pm5\,\%$. Threshold for recovering the correlation peak is significance of 7, providing a probability of $\approx10^{-12}$ of a wrongly found peak \cite{Ho.2009}. clock frequency skew compensation increases the visibility of the correlation peak and thus enables higher noise resistance in this work. \textbf{b} Coincidence-to-accidentals ratio depending on the transmission from Alice to Bob (Methods \ref{SI:ho_vs_our_work}, equation \ref{eq:CAR_equation}). }
	\label{fig:SI_significance_vs_SNR}
\end{figure*}

More detailed relations between significance and the signal-to-noise ratio are depicted in Fig. \ref{fig:SI_significance_vs_SNR}. Starting from single count rates at Alice $r_A$ and Bob's side $r_B$, the single rate at Bob's reduces as the link transmission $T$ reduces by means of a variable attenuator. Furthermore reduces the coincidence rate $r_C$ similarly. Following Refs. \cite{Zhao.2020,He.2015,C.Xiong.2011}, the coincidence-to-accidentals ratio ($CAR$) is defined as the number of true coincidence counts $r_\text{CTrue}$ over the accidental coincidence counts $r_\text{CAcc}$,

\begin{equation}
    CAR = \dfrac{r_\text{CTrue}}{r_\text{CAcc}} = \dfrac{r_C-r_\text{CAcc}}{r_\text{CAcc}} .
\end{equation}

\noindent
The accidental correlation events, depending on the link transmission $T$, within the root-mean-squared coincidence window $\sigma$ is,

\begin{equation}
    r_\text{CAcc}(T) = 2 r_A \left[ (r_B-r_\text{back})T + r_\text{back}\right] \sigma .
\end{equation}

\noindent
Here we also introduce the background rate $r_\text{back}$ that can not be further reduced by higher channel losses. This includes factors like detector dark counts, not sufficiently filtered daylight, or other other noise sources in the system. The transmission-dependent $CAR$ summarizes as 

\begin{equation}
    CAR(T) = \dfrac{r_CT-2 r_A \left[ (r_B-r_\text{back})T + r_\text{back}\right] \sigma}{2 r_A \left[ (r_B-r_\text{back})T + r_\text{back}\right] \sigma} . \label{eq:CAR_equation}
\end{equation}

\noindent
The transmission-dependent peak significance can be derived from equation \ref{eq:ho_S_p} as, 

\begin{equation}
    S_p(T) = \sqrt{\dfrac{(r_CT)^2}{r_A \left[ (r_B-r_\text{back})T + r_\text{back}\right] \Delta u}} . \label{eq:analytical_Sp_T}
\end{equation}

\noindent
The analytical trend, as described by equation \ref{eq:analytical_Sp_T}, has been confirmed experimentally (Fig. \ref{fig:SI_significance_vs_SNR}) after compensation of the clock frequency skew by different amounts. The peak significance is achieved by matching the cross-correlation bin size with the peak spread from the residual clock frequency skew, as described earlier. However, for clock frequency skews smaller than 10\,ns/s is the number of bins already $N = 1/ (10\times 10^{-9}) = 10^8$ that results in immense computation effort (more information in Methods \ref{SI:cross_correlation_computation_and_algorithm}). It is hardly feasible to adapt the bin size to its optimum for clock frequency skews of $\leq10\,ns/s$. It follows a non-optimized cross-correlation that does not provide maximum significance. Nevertheless, Fig. \ref{fig:SI_significance_vs_SNR} shows impressively how the significance of correlation peaks can be increased by compensation of the clock frequency skew.

\begin{figure*}[htb]
	\centering
    \includegraphics[width=71.086mm]{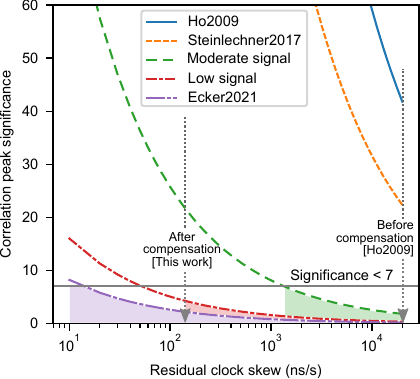}
	\caption{\textbf{Correlation peak significance for various experimental setups} The clock frequency skew of quartz oscillator is up to 20\,$\mu$s/s, providing the clock frequency skew before compensation. Even without compensation it is easy to find the correlation peaks with sufficient signal, as in Ho2009 \cite{Ho.2009} and Steinlechner2017 \cite{Steinlechner.2017}. Threshold for recovering the correlation peak is significance of 7, giving a probability of $\approx10^{-12}$ of a wrongly found peak \cite{Ho.2009}. Low coincidence-to-accidentals ratios, as in our low signal (Methods \ref{SI:high_loss_live_tracking} and Table \ref{tab:table1}a) or moderate signal experiment (Fig. \ref{fig:figure_3} and Table \ref{tab:table1}b) or in Ecker2021 \cite{Ecker.2021}, require compensation of the clock frequency skew. Under our limited computation power we can reduce and compensate the clock frequency skew to $<140$\,ns/s. This is not sufficient to reach a significance of 7 in our low signal experiment or in Ecker2021 \cite{Ecker.2021}. Solution would be to reduce the initial clock frequency skew search window.}
	\label{fig:HoSteinEcker_vs_us}
\end{figure*}

Reduction of the clock frequency skew by our algorithm can help to recover the correlation features under very low signal-to-noise ratios. High signal rates do not demand any kind of compensation. On the other hand, it is crucial in high-loss link scenarios (Fig. \ref{fig:HoSteinEcker_vs_us} with data from table \ref{tab:table1}). Here we are limited to a clock frequency skew compensation to 140\,ns/s that allows for synchronization initialization of our emulated link experiment (Fig. \ref{fig:figure_3} and table \ref{tab:table1}b). Even lower rates, as in Methods \ref{SI:high_loss_live_tracking} (table \ref{tab:table1}a) or in \cite{Ecker.2021} causes too small correlation peak significance for synchronization initialization today (Fig. \ref{fig:HoSteinEcker_vs_us}). Higher correlation peak significance is expected to reach with more computation power or reduction of the clock frequency skew search window.

\newpage
\subsection{Synchronization initialization algorithms and computation efforts}\label{SI:cross_correlation_computation_and_algorithm}

\begin{figure}[htb!]
\begin{minipage}{1\linewidth}
  \begin{algorithm}[H]
\begin{align*}
\textbf{Input: }& \text{timetags Alice } t_A \text{ and Bob } t_B\\
&\text{Functions: read timetags(), convolution(),} \\
\textbf{Output: }& \text{optimum clock frequency skew}
\end{align*}

$\text{sleep(integration time = 0.1 sec) } $
    
$t_A,t_B = \text{read timetags}()$

$\text{skew vector} = -20\,\mu s/s \dots +20\,\mu s/s, \, \text{step: } 140\,ns/s$

$\text{bins} = 0 \dots 0.1 s,\, \text{step: } 14\,ns$

$P = \text{zeros}(\text{skew vector size})$

$i = 0$

\For{$\Delta u_{\text{corr}}$ \textbf{in} \text{skew vector}}{
    $t_{B_{\text{corr}}} = t_B + (t_B - t_B[0]) \times \Delta u_{\text{corr}}$
    
    $ c = \text{convolution}(t_A,t_{B_{\text{corr}}},\text{bins})$

    $P[i] = \text{max}( c )$
    
    $i +=  1$
    }

$\text{optimum clock frequency skew} = \text{skew vector}( \text{argmax}(P) ) $

  \end{algorithm}
\caption[]{\textbf{Representation of the function coarse clock frequency skew() and accurate clock frequency skew() in Methods \ref{SI:live_tracking_algorith}, Fig. \ref{fig:SI_Rb_clock_skew_compensation}} The convolution is fast Fourier transform (FFT)-based and Start-Stop method-based, respectively. A FFT-based approach has to be taken if the precise timing offset is not known. The algorithm includes the following variables, time tags from Alice and Bob $t_{A/B}$, peak values of the cross-correlations $P$, counter $i$, clock frequency skew for correction $\Delta u_\text{corr}$, clock-skew correct time tags from Bob $t_{B_{\text{corr}}}$, cross-correlation output $c$. Note that the bin step size of 14\,ns corresponds to integration time $\times$ skew vector step size. Smaller bin and skew vector step sizes require considerably higher computation effort. If there is knowledge of the approximate clock frequency skew, the range of clock frequency skews can be reduced.}\label{alg:clock_skew}
\end{minipage}
\end{figure}

The correlation peak is found by a convolution of the time tag stream from Alice and Bob. Without precise knowledge of the clock frequency skew, it will be difficult to find correlation peaks with sufficient significance. First, the clock frequency skew search window is created (Fig. \ref{alg:clock_skew}). From our quartz oscillator data sheet, we know the approximate range of clock frequency skews, starting from -20 to +20\,$\mu$s/s (see Methods \ref{SI:experimental_clock_drift}). The time tags from Bob are compensated by different clock frequency skews every loop analogue to equation \ref{eq:skew_correction} and then convoluted with time tags from Alice. It is either based on fast Fourier transform, in the case of determining the clock frequency skew coarsely (Fig. \ref{fig:figure_2}a) without timing offset knowledge, or it is based on the Start-Stop method. In the case of fast Fourier transform, it is required to arrange the time tags to the bins and then apply the cross-correlation. In the Start-Stop method are time differences binned after the correlation. 

\begin{table*}[htb]
\caption{\label{tab:personal_computer}Personal computer specifications for calculating the fast Fourier transform.}
\begin{tabular}{ |c|c|  }
\hline
Parameter& Value\\
\hline
Processor & Intel(R) Core(TM) i5-8250U CPU \\
Speed &  1.60\,GHz (1.80\,GHz)\\
RAM & 16\,GB \\
System & 64-bit based processor \\
Environment & Python 3.7.0 64\,bit, NumPy 1.19.4 \\
\hline
\end{tabular}
\end{table*}

\begin{figure}[htb]
	\centering
	\includegraphics[width=70.819mm]{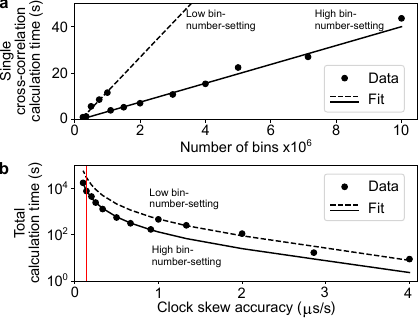}
	\caption{\textbf{Fast Fourier transform computation times for different clock frequency skew accuracy.} \textbf{a} Single-cross-correlation computation time depending on the number of bins. The personal computer (Table \ref{tab:personal_computer}) uses two different settings for calculations, depending on the number of bins. In case of adaption of the bin width to the correlation peak spread is the clock frequency skew accuracy equal to the inverse of the number of bins (equation \ref{eq:clock_frequency_accuracy_number_of_bins}). \textbf{b} Total computation time for a search window of $-20\,\mu$s/s to $+20\,\mu$s/s with given clock frequency skew accuracy (step size).}
	\label{fig:computation_effort}
\end{figure}

Full cross-correlations over all time tags are time-consuming and limit the feasible number of calculations. The correlations are performed through a Python environment with the NumPy module on the personal computer (Table \ref{tab:personal_computer}). The time for a single fast Fourier transform-based cross-correlation may amount up to approximately 40 seconds for $N = 10^7$ bins (Fig. \ref{fig:computation_effort}a). Optimum significance is reached by choosing enough number of bins $N$, so that the correlation peak spread is equal to the bin size (see Methods \ref{SI:ho_vs_our_work}). This gives the achievable clock frequency skew accuracy $\delta u$,

\begin{equation}
    \delta u = \dfrac {1}{N}.\label{eq:clock_frequency_accuracy_number_of_bins}
\end{equation}

The total computation time $T_{\text{tot}}$ from a single cross-correlation $T(\delta u)$ for a range of clock frequency skews $\Delta U$ is then,

\begin{equation}
    T_{\text{tot}} = \dfrac{\Delta U}{\delta u}\times T(\delta u).
\end{equation}

With a time limit of slightly above 2 hours (7500 seconds) and a clock frequency skew range of $\Delta U = 40\,\mu$s/s (from -20 to +20\,$\mu$s/s) is the maximum affordable time for a single cross-correlation approximately 26 seconds and clock frequency skew accuracy $0.14\,\mu$s/s ($N = 7.14\times10^6$). The speed of 26 seconds for a single cross-correlation may be drastically increased by using a dedicated higher-performing computer, instead of a laptop here. More specifically, more processing cores and higher processing speed could improve the situation. Graphical processing units may also perform much better by parallelization of the processes and for-loops.

\subsection{Estimation of synchronization jitter and trend lines}

\subsubsection{Description of total system jitter} \label{SI:total_system_jitter}

The total system jitter of a quantum communication system determines the overall performance - from signal to noise ratio to secure bit rate. The total jitter can be easily calculated analytically by knowledge of the clock frequency skew. Here we provide important scaling laws to estimate the final signal-to-noise ratios, like significance or coincidence to accidentals ratio. As all jitter contributions originate from a random source, we consider a Gaussian function $g$ with RMS jitter $\sigma$ that is normalized to have an area of 1,
\begin{equation}
    g(t) = \dfrac{1}{\sqrt{2\pi\sigma^2}}\exp\left(-\frac{t^2}{2\sigma^2}\right). \label{eq:Gaussian}
\end{equation}

\noindent
The total jitter from the source coherence time $\sigma_{\text{coh}}$, time tagger $\sigma_{\text{tt}}$, detector $\sigma_{\text{det}}$ and synchronization $\sigma_{\text{sync}}$,
\begin{equation}
    \sigma = \sqrt{ \sigma_{\text{coh}}^2 + \sigma_{\text{tt}}^2 + \sigma_{\text{det}}^2 + \sigma_{\text{sync}}^2 }, 
\end{equation}

\noindent
reduces to,
\begin{equation}
    \sigma \approx \sqrt{\sigma_{\text{det}}^2 + \sigma_{\text{sync}}^2 }, \label{eq:sigma_short}
\end{equation}

\noindent
as the jitter contribution from the time tagger (around 30\,ps root-mean-squared) and entangled photon source coherence time (around 2.5\,ps root-mean-squared) is much smaller than the detector jitter. After a few steps of editing, we get
\begin{equation}
    \sigma = \sigma_{\text{det}}\sqrt{1 + \dfrac{\sigma_{\text{sync}}^2}{\sigma_{\text{det}}^2} }, \label{eq:total_sigma}
\end{equation}

\noindent
resulting in the simulated jitter curves in Fig. \ref{fig:figure_2}a. Due to the residual clock frequency skew after compensation is the smallest jitter not equal to the detector jitter.  The highest correlation peak value $P$ (replaceable by coincidence to accidentals-ratio $CAR$ or significance $S$), given by the smallest timing jitter $\sigma_{\text{min}}$, reduces, due to the stronger synchronization jitter. With equation \ref{eq:sigma_short} is the correlation peak value described as,
\begin{equation}
    P = \dfrac{P_0}{ \sqrt{1+\left(\dfrac{2\sigma_\text{sync}}{\sqrt{2\pi}\sigma_\text{min}}\right)^2 }}. \label{eq:peak_significance_vs_skew}
\end{equation}

\begin{figure}[htb]
	\centering
	\includegraphics[width=77.373mm]{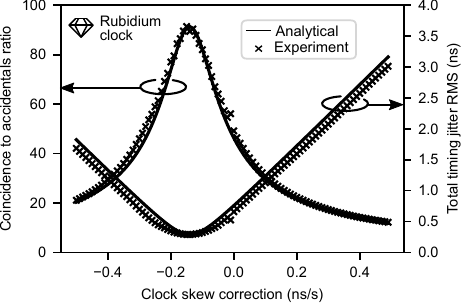}
	\caption{Coincidence-to-accidentals ratio and total timing jitter for corrected clock frequency skew of rubidium oscillators. The smooth trend indicates a stable reference. The analytical trend lines are derived from the experimental acquisition time of $T_a = 10\,$seconds, smallest timing jitter of 320\,ps and maximum coincidence-to-accidentals ratio of 91.}
	\label{fig:SI_RbClockCAR}
\end{figure}

\noindent
The prefactor of $\sqrt{2/\pi}$ originates from the area of a Gaussian being equal to $\sqrt{2\pi}\sigma$ and its changing shape to a flat-top Gaussian by the synchronization jitter. Figure \ref{fig:SI_RbClockCAR} indicates a smooth trend of $CAR$ and total timing jitter RMS for rubidium clocks. The smallest experimental timing jitter amounts to $\sigma_\text{min} = 310\,$ps with maximum coincidence-to-accidentals ratio $CAR_0$ of 91 and acquisition time $T_a$ of 10\,seconds. Both together enable an analytical trend line from equations \ref{eq:sync_jitter},\ref{eq:total_sigma} and \ref{eq:peak_significance_vs_skew} with acquisition time $T_a = 10\,$s (Fig. \ref{fig:SI_RbClockCAR}),

\begin{equation}
    CAR(\Delta u) = \dfrac{CAR_0}{ \sqrt{1+\dfrac{2}{\pi}\left(\dfrac{1/2\Delta u T_a}{\sigma_\text{min}} \right)^2 } },
\end{equation}

\begin{equation}
    \sigma(\Delta u) = \sigma_\text{min} \sqrt{1 + \left(\dfrac{1/2 \Delta u T_a}{\sigma_\text{min}} \right)^2 } . \label{eq:trend_sigma_delta} 
\end{equation}

\noindent
The equation \ref{eq:peak_significance_vs_skew} may also be transformed to be dependent on the residual clock frequency skew $\Delta u$ and smallest residual clock frequency skew $\delta u$, as the smallest timing jitter $\sigma_\text{min}$ and smallest residual clock frequency skew $\delta u$ are directly related through equation \ref{eq:sync_jitter},

\begin{equation}
    S(\Delta u, S_p) = \dfrac{S_p}{ \sqrt{1+\dfrac{2}{\pi}\left(\dfrac{\Delta u}{\delta u (S_p)} \right)^2 }}, \label{eq:trend_Sp_deltau}
\end{equation}

\noindent
with the smallest residual clock frequency skew $\delta u$ defined by\cite{Ho.2009},

\begin{equation}
    \delta u (S_p) = \dfrac{r_C^2}{r_A r_B S_p^2}. \label{eq:clock_drift_max}
\end{equation}

\noindent

The maximum significance of 142 allows to create helpful trend lines with the coincidence rate $r_C= 7$\,kcps, single rate on Alice's side $r_A = 244$\,kcps, Bob's side 2$r_B = 232$\,kcps and acquisition time of 0.1\,seconds (see Methods \ref{SI:total_system_jitter}). The analytical trend of the significance directly derives from the maximum experimental significance $S_p$ and follows equation \ref{eq:trend_Sp_deltau} in Methods \ref{SI:total_system_jitter}. The smallest residual clock frequency skew at the peak significance can be calculated simultaneously to $\delta u = 48\,$ns/s with $S_p = 142$ (see Methods \ref{SI:total_system_jitter} equation \ref{eq:clock_drift_max}), and leads to the smallest timing jitter $\sigma_\text{min} = 2.4$\,ns with equation \ref{eq:sync_jitter}. Furthermore, the peak significance $S_p$ enables analytical trend lines of the total timing jitter $\sigma$ as $\sigma^2 = \sigma_\text{min}^2 + \sigma_\text{sync}^2$, providing a close match with the experimental timing jitters (Fig. \ref{fig:figure_2}a). 

\subsubsection{Achievable synchronization timing jitter during live tracking}\label{SI:live_correction_jitter}

The final synchronization jitter after live correction during a communication session is decided by the clock drift and the number of correlation events. Here we present analytical estimations of the synchronization limits that might be helpful for easy transfer to any communication scenario. The most important parameter is the instantaneous clock frequency skew during a communication session. If it is nonzero, due to insufficient correction, the synchronization jitter will be nonzero. The clock frequency skew $\Delta u$ is calculated from two or more detected peak locations $\tau_1$ and $\tau_2$ with temporal separation of $T_\text{meas}$,
\begin{equation}
    \Delta u = \dfrac{\tau_2-\tau_1}{T_\text{meas}}.
\end{equation} 
However, low numbers of $n$ correlation events give rise to an uncertainty of the correlation peak location $\delta \tau$ via the total timing jitter $\sigma$ \cite{Ho.2009},
\begin{equation}
    \delta\tau = \dfrac{\sigma}{\sqrt{n-1}}. \label{eq:location_error}
\end{equation}
Via error propagation, we get the uncertainty of the clock frequency skew from measurement $\Delta u_\text{meas}$,
\begin{equation}
    \Delta u_\text{meas} = \sqrt{2}\dfrac{\delta\tau}{T_\text{meas}}.
\end{equation}
Together with equations \ref{eq:sync_jitter} and \ref{eq:location_error}, coincidence rate $r_C$ and acquisition time $T_a$, is the synchronization jitter due to the uncertainty of peak position measurement,
\begin{equation}
    \sigma_{\text{meas}} = \frac{1}{2} \sqrt{2}\dfrac{\sigma}{ \sqrt{ r_C T_a - 1 } T_\text{meas}} T_a.\label{eq:sigma_meas}
\end{equation}

The second jitter contribution comes from the clock drift $\partial (\Delta u) / \partial t$. clock frequency skew, that has not been foreseen previously, may accumulate over the feedback loop time $T_{\text{feed}}$,
\begin{equation}
     \Delta u_\text{drift} = \dfrac{\partial (\Delta u)}{\partial t} T_\text{feed}.
\end{equation}
Together with equation \ref{eq:sync_jitter} is the resulting synchronization jitter, due to the drifting clock,
\begin{equation}
    \sigma_{\text{drift}} = \frac{1}{2} \dfrac{\partial (\Delta u)}{\partial t} T_\text{feed} T_a. 
\end{equation}
Both measurement and synchronization jitter limit the total achievable synchronization jitter during live tracking (Fig. \ref{fig:figure_2}c), 
\begin{equation}
    \sigma_\text{sync}^2 = \sigma_{\text{meas}}^2 +  \sigma_{\text{drift}}^2.
\end{equation}

\begin{figure*}[htb]
	\centering
	\includegraphics[width=1\linewidth]{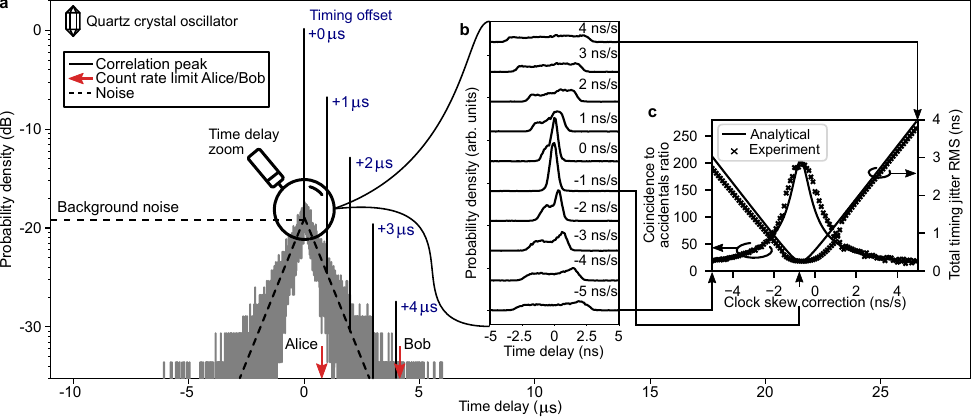}
	\caption[Second Figure]{\textbf{Fine tuning of clock frequency skew through cross-correlation with Start-Stop method} \textbf{a} Correlation peaks for various timing offset accuracy. The noise values are highest at small time delay with -20 dB and reduce exponentially until approximately the inverse count rate at Bob is reached. The peak noise level is larger than the correlation peak for poor timing offset precision $> 2.5\,\mu$s. Peak search algorithms may still recover the peak here, instead of simple search of maximum at timing offset precision $< 2.5\,\mu$s. \textbf{b} Magnified view of timing offset-corrected cross-correlation peaks for various clock frequency skew compensation values. The shape of correlation peaks depends on the compensated clock frequency skew and may be asymmetrical from clock instabilities. \textbf{c} The correlation peak coincidence to accidentals ratio or its timing jitter is ideal indicator for optimum clock frequency skew compensation, as they become maximum and minimum, respectively. The analytical trend lines are derived from the experimental acquisition time of 1.4\,seconds, smallest timing jitter of 258\,ps and maximum coincidence-to-accidentals ratio of 204.}
	\label{fig:clock_skew_finetuning}
\end{figure*}
\newpage

\subsection{Clock frequency skew fine tuning and Start-Stop method} \label{SI:clock_skew_fine_tuning}

Careful fine tuning of the optimum clock frequency skew for compensation reduces the synchronization jitter to a minimum. As the timing offset is calculated with nanosecond accuracy in the synchronization initialization, it is feasible to shrink down the observation window to $<100\,$ns to reduce the computation effort. Cross-correlations are now derived by the computation-efficient Start-Stop method \cite{Brunel.1999,Alleaume.2004,Martinez.2016}. In contrast to the FFT cross-correlation that correlates all timetags from Alice with all Bob timetags from Bob, the Start-Stop method calculates time differences between neighboring timetags. If the timing offset was not calculated with a precision smaller than $\text{Min}\{1/r_A,1/r_B\}$, it rises the probability of wrongly calculated time differences and thus reduces the number of correlation events. The correlation peak height reduces and finally becomes smaller than the peak noise values (Fig. \ref{fig:figure_2}a). The correlation peak compresses to the smallest width $\sigma_{\text{min}} = 258\,$ps by compensation of the residual clock frequency skew. In addition maximizes the coincidence-to-accidentals ratio at $\text{CAR}_0 = 204$ over a coincidence window that is determined by the total timing jitter. The experimental correlation peak is fit by a stretched Gaussian function and then the timing jitter derived from it. The experimental total jitter and peak value behaves perfectly as analytically given (see Methods \ref{SI:total_system_jitter}). Clocks with high stability, i.e. a constant frequency difference between two clocks, create a jitter envelope with a plateau. Rubidum oscillators are known for their high stability of $10^{-12}$ over 1 second \cite{Vanier.1981,Penrod.1996} and create a jitter envelope with a plateau, as the frequency difference is almost constant over time. Quartz crystal oscillators on the other hand may show weak stability of only $10^{-11}$ to $10^{-9}$ over 1 second \cite{TiradoAndres.2019} that result in asymmetry and give rise to deviations between analytical and experimental trends in a region between -1 to 2\,ns/s. The method for compensation of the clock frequency skew by a simple peak search (Fig. \ref{fig:figure_2}c) may be done with arbitrary resolution. Step sizes of $<1\,$ns/s work greatly to start live tracking of clock drifts in the last step of synchronization. 

\begin{figure*}[htb]
	\centering
	\includegraphics[width=0.95\linewidth]{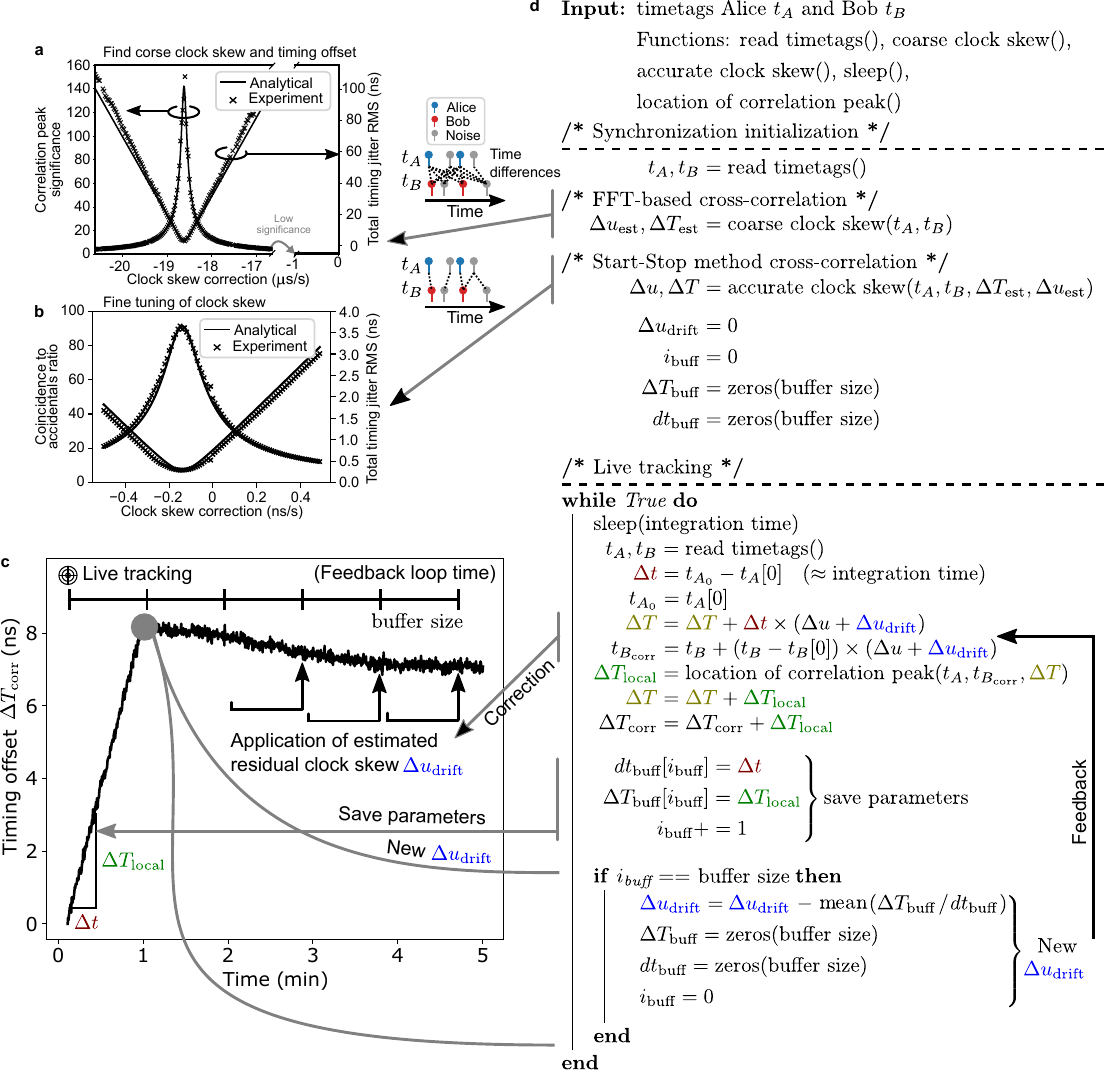}
	\caption[Second Figure]{\textbf{Initialisation and live tacking algorithm} \textbf{a} Coarse search of the clock frequency skew. \textbf{b} Fine-tuning of the clock frequency skew. \textbf{c} Estimation of residual clock frequency skew over a feedback time of approximately 1 minute with two rubidium clocks. The residual clock frequency skew over the last 4 minutes amounts to 6.4\,ps/s, resulting in synchronization jitters of approximately 0.32\,ps over acquisition time of 100\,ms. \textbf{d} The algorithm for synchronization initialization and live tracking includes the the following variables, time tags from Alice and Bob $t_{A/B}$, first estimated clock frequency skew $\Delta u _\text{est}$, estimated timing offset $\Delta T_\text{est}$, accurate timing offset $\Delta T$ and clock frequency skew $\Delta u$, instantaneous clock frequency skew from clock drifts $\Delta u_{\text{drift}}$, buffer of local timing offset $\Delta T_\text{buff}$, buffer of local time difference $dt_\text{buff}$, buffer counter $i_\text{buff}$, time difference $\Delta t$, time of previous sample $t_{A_0}$, clock-skew corrected time tags from Bob $t_{B_{\text{corr}}}$, local timing offset $\Delta T_\text{local}$. The location of the correlation peak is found through an efficient Start-Stop cross-correlation.}
	\label{fig:SI_Rb_clock_skew_compensation}
\end{figure*}
\newpage

\subsection{Live tracking algorithm} \label{SI:live_tracking_algorith}

After synchronization initialization, including the coarse clock frequency skew search (Fig. \ref{fig:SI_Rb_clock_skew_compensation}a) and its fine-tuning (Fig. \ref{fig:SI_Rb_clock_skew_compensation}b), residual and time-dependent clock frequency skews increase the synchronization jitter and call for compensation during communication sessions. Fig. \ref{fig:SI_Rb_clock_skew_compensation}c represents the effect of correct compensation for rubidium clocks by the algorithm in Fig. \ref{fig:SI_Rb_clock_skew_compensation}d. The correlation peak timing offset is tracked over $n$ data points. Any change of timing offset would be caused by an instantaneous drift. Here, the timing offset changes by approximately 8\,ns in 1 minute. By applying the measured residual clock frequency skew to future time tags, we can reduce the timing offset change and subsequently the synchronization jitter. Long averaging times work especially well for highly stable clocks, such as rubidium clocks. The timing offset changes by less than  1.5\,ns over 4 minutes after the first correction, already providing residual clock frequency skews of $<6.4\,$ps/s and thus easily synchronization jitters $<0.32\,$ps in typical 100\,ms data package sizes.

\subsection{Photon pair rates} \label{SI:exp_photon_pair_rates}
\begin{figure}[htb]
	\centering
	\includegraphics[width=70.686mm]{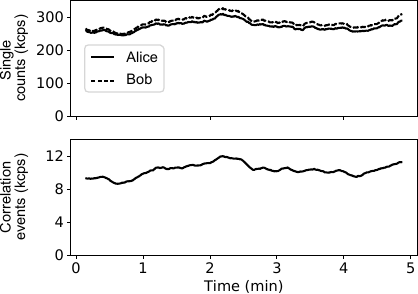}
	\caption{Photon pair rates according to Fig. \ref{fig:figure_2}b} \label{fig:SI_pair_rate}
\end{figure}

The experimentally detected photon pair rates determine achievable signal-to-noise ratios and limit the synchronization jitter under high losses. Figure \ref{fig:SI_pair_rate} represents the rates from experiment (Fig. \ref{fig:figure_2}b). 

\newpage
\subsection{Experimental clock drift} \label{SI:experimental_clock_drift}
\begin{figure}[htb]
	\centering
	\includegraphics[width=68.4mm]{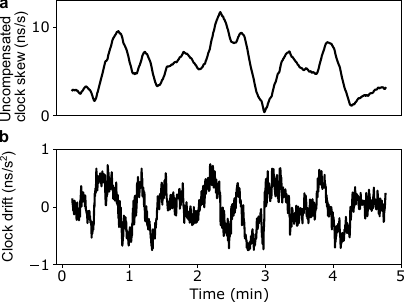}
	\caption[Second Figure]{\textbf{a} The instantaneous clock frequency skew $\Delta u$ drifts without live tracking. \textbf{b} The time-dependent clock drift $\partial (\Delta u)/\partial t$ is zero on average with standard deviation of 320\,ps/s$^2$.}
	\label{fig:SI_clock_drift}
\end{figure}

Knowledge of the clock stability provides important measures about the limiting synchronization jitters. To estimate the clock stability, we first compare the timing jitter without tracking to the reference (Fig. \ref{fig:figure_3}b). Any increase from the reference is caused by synchronization jitters through the residual clock frequency skews $\Delta u$ (equation \ref{eq:sync_jitter}). The residual, not compensated clock frequency skew is plotted in Fig. \ref{fig:SI_clock_drift}a and amounts to more than 10\,ns/s. Almost ideal synchronization may be achieved if the residual clock frequency skew is constant. However, crystal oscillators may have time-dependent frequencies that result in variation of the residual clock frequency skew. Fig. \ref{fig:SI_clock_drift}b indicates the acceleration of the clocks with time with a mean value of zero and standard deviation of $320\,$ps/s$^2$. High clock drifts require small feedback times for compensation that simultaneously demands many correlation events.  
\newline

The time taggers are equipped with external crystal oscillators without temperature control. The data sheet accuracy is $\pm 20\,$ppm, aging $\pm 3\,$ppm/ first year, $\pm 1\,$ppm/year and temperature dependence $\pm 0.125\,$ppm (25°C ... 85°C). The rubidium clocks are temperature controlled with accuracy $\pm10^{-4}$\,ppm (ambient temperature 0°C ... 40°C), aging $<5\times10^{-5}$\,ppm/month, stability over 1s is 10$^{-5}$\,ppm. 

\subsection{Low signal live tracking}\label{SI:high_loss_live_tracking}

\begin{figure*}[htb]
	\centering
    \includegraphics[width=170mm]{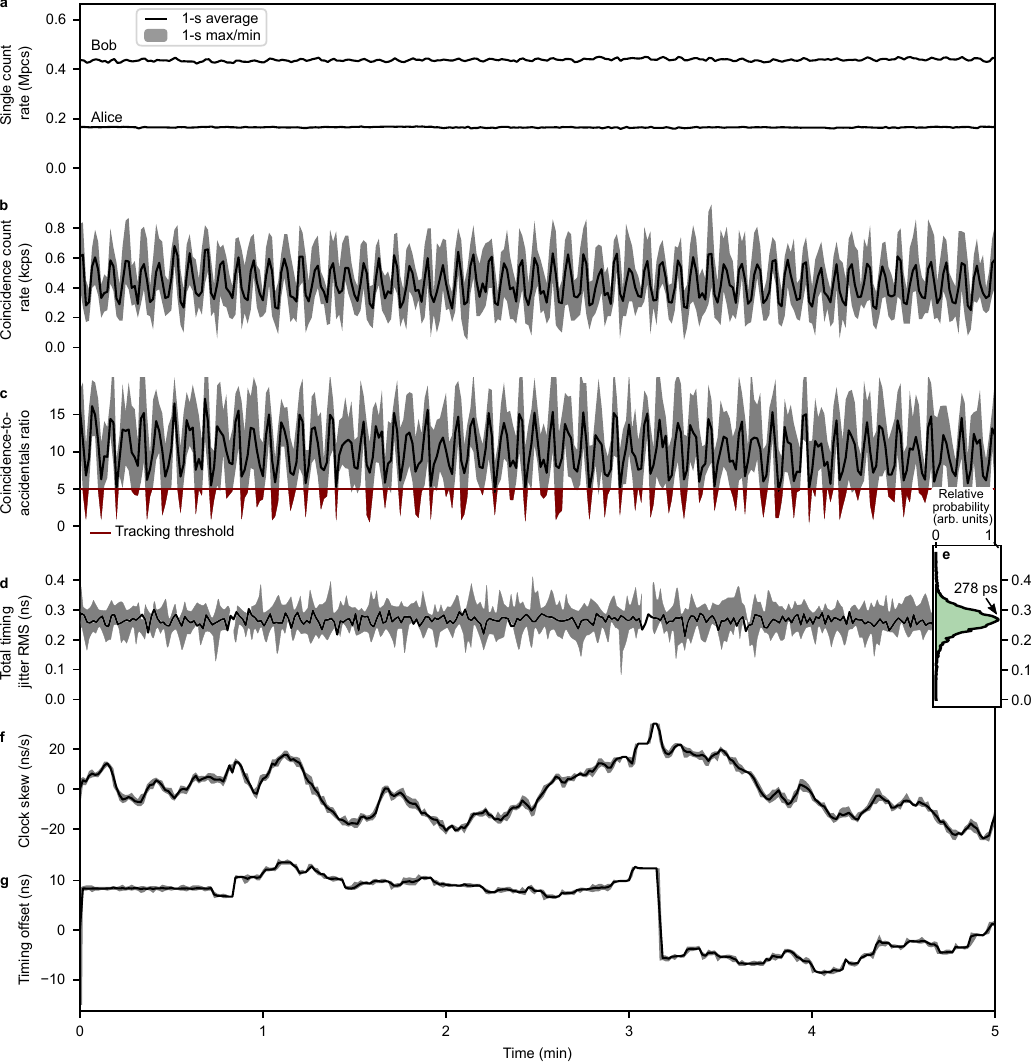}
	\caption{\textbf{Low correlation rate live tracking under turbulence.} Turbulence parameter: Fried parameter 1\,mm, $1/e^2$ full-width beam waist of 3.4\,mm. \textbf{a} The count rate at the receiver Bob of $437\pm6$\,kcps is increased artificially by a noise source. Alice's count rate amounts to $165\pm3$\,kcps. \textbf{b} The average coincidence rate is $430\pm160$\,cps (coincidence window is equal to RMS timing jitter of 267\,ps) and indicates periodic fluctuations from the turbulence disk. \textbf{c} The mean coincidence-to-accidentals ratio is $10\pm4$ with a correlation peak live tracking threshold of 5. \textbf{d-e} The total timing jitter probability distribution describes a Gaussian with its fitted center at $277.8\pm0.8$\,ps. Together with same-clock total timing jitter of $260\pm2$ps is the synchronization jitter approximately $98\pm6$\,ps.
	\textbf{f} The tracked clock frequency skew displayed over time.
	\textbf{g} The timing offset varies more than in high-signal environment (Fig. \ref{fig:figure_3}), because of partially stopped tracking when the coincidence-to-accidentals ratio falls below 5. This gives rise to an increase of the total timing jitter. The feedback loop time and data package size amounts to 200\,ms and 100\,ms, respectively. Summary of this figure in table \ref{tab:table1}.}
	\label{fig:low_SNR_tracking}
\end{figure*}

\newpage


\bibliography{main}

\providecommand{\noopsort}[1]{}\providecommand{\singleletter}[1]{#1}%
\begin{thebibliography}{10}

\bibitem{Bellamy.1995}
J.~C. Bellamy.
\newblock Digital network synchronization.
\newblock {\em IEEE Communications Magazine}, 33(4):70--83, 1995.

\bibitem{Narula.2018}
Lakshay Narula and Todd~E. Humphreys.
\newblock Requirements for secure clock synchronization.
\newblock {\em IEEE Journal of Selected Topics in Signal Processing},
  12(4):749--762, 2018.

\bibitem{Phadke.1994}
A.~G. Phadke, B.~Pickett, M.~Adamiak, M.~Begovic, G.~Benmouyal, R.~O. Burnett,
  T.~W. Cease, J.~Goossens, D.~J. Hansen, M.~Kezunovic, L.~L. Mankoff, P.~G.
  McLaren, G.~Michel, R.~J. Murphy, J.~Nordstrom, M.~S. Sachdev, H.~S. Smith,
  J.~S. Thorp, M.~Trotignon, T.~C. Wang, and M.~A. Xavier.
\newblock Synchronized sampling and phasor measurements for relaying and
  control.
\newblock {\em IEEE Transactions on Power Delivery}, 9(1):442--452, 1994.

\bibitem{Angel*.2014}
James~J. Angel*.
\newblock When finance meets physics: The impact of the speed of light on
  financial markets and their regulation.
\newblock {\em Financial Review}, 49(2):271--281, 2014.

\bibitem{Corbett.2013}
James~C. Corbett, Jeffrey Dean, Michael Epstein, Andrew Fikes, Christopher
  Frost, J.~J. Furman, Sanjay Ghemawat, Andrey Gubarev, Christopher Heiser,
  Peter Hochschild, Wilson Hsieh, Sebastian Kanthak, Eugene Kogan, Hongyi Li,
  Alexander Lloyd, Sergey Melnik, David Mwaura, David Nagle, Sean Quinlan,
  Rajesh Rao, Lindsay Rolig, Yasushi Saito, Michal Szymaniak, Christopher
  Taylor, Ruth Wang, and Dale Woodford.
\newblock Spanner: Google's globally distributed database.
\newblock {\em ACM Trans. Comput. Syst.}, 31(3), 2013.

\bibitem{Bennett.2014}
Charles~H. Bennett and Gilles Brassard.
\newblock Quantum cryptography: Public key distribution and coin tossing.
\newblock {\em Theoretical Computer Science}, 560:7--11, 2014.

\bibitem{Ekert.1991}
Artur~K. Ekert.
\newblock Quantum cryptography based on bell's theorem.
\newblock {\em Physical review letters}, 67(6):661--663, 1991.

\bibitem{Northup.2014}
T.~E. Northup and R.~Blatt.
\newblock Quantum information transfer using photons.
\newblock {\em Nature Photonics}, 8(5):356--363, 2014.

\bibitem{Gisin.2007}
Nicolas Gisin and Rob Thew.
\newblock Quantum communication.
\newblock {\em Nature Photonics}, 1(3):165--171, 2007.

\bibitem{Chen.2021}
Yu-Ao Chen, Qiang Zhang, Teng-Yun Chen, Wen-Qi Cai, Sheng-Kai Liao, Jun Zhang,
  Kai Chen, Juan Yin, Ji-Gang Ren, Zhu Chen, Sheng-Long Han, Qing Yu, Ken
  Liang, Fei Zhou, Xiao Yuan, Mei-Sheng Zhao, Tian-Yin Wang, Xiao Jiang, Liang
  Zhang, Wei-Yue Liu, Yang Li, Qi~Shen, Yuan Cao, Chao-Yang Lu, Rong Shu,
  Jian-Yu Wang, Li~Li, Nai-Le Liu, Feihu Xu, Xiang-Bin Wang, Cheng-Zhi Peng,
  and Jian-Wei Pan.
\newblock An integrated space-to-ground quantum communication network over
  4,600 kilometres.
\newblock {\em Nature}, 589:214--219, 2021.

\bibitem{Mills.2011}
David~L. Mills.
\newblock {\em Computer network time synchronization: The Network Time Protocol
  on Earth and in space}.
\newblock {CRC Press}, Boca Raton FL, 2nd ed. edition, 2011.

\bibitem{Berceau.2016}
P.~Berceau, M.~Taylor, J.~Kahn, and L.~Hollberg.
\newblock Space-time reference with an optical link.
\newblock {\em Classical and Quantum Gravity}, 33(13):135007, 2016.

\bibitem{Tomita.2010}
Akihisa Tomita, Ken-ichiro Yoshino, Yoshihiro Nambu, Akio Tajima, Akihiro
  Tanaka, Seigo Takahashi, Wakako Maeda, Shigehito Miki, Zhen Wang, Mikio
  Fujiwara, and Masahide Sasaki.
\newblock High speed quantum key distribution system.
\newblock {\em Optical Fiber Technology}, 16(1):55--62, 2010.

\bibitem{Yin.2020}
Juan Yin, Yu-Huai Li, Sheng-Kai Liao, Meng Yang, Yuan Cao, Liang Zhang, Ji-Gang
  Ren, Wen-Qi Cai, Wei-Yue Liu, Shuang-Lin Li, Rong Shu, Yong-Mei Huang, Lei
  Deng, Li~Li, Qiang Zhang, Nai-Le Liu, Yu-Ao Chen, Chao-Yang Lu, Xiang-Bin
  Wang, Feihu Xu, Jian-Yu Wang, Cheng-Zhi Peng, Artur~K. Ekert, and Jian-Wei
  Pan.
\newblock Entanglement-based secure quantum cryptography over 1,120 kilometres.
\newblock {\em Nature}, 582(7813):501--505, 2020.

\bibitem{Liao.2017}
Sheng-Kai Liao, Wen-Qi Cai, Wei-Yue Liu, Liang Zhang, Yang Li, Ji-Gang Ren,
  Juan Yin, Qi~Shen, Yuan Cao, Zheng-Ping Li, Feng-Zhi Li, Xia-Wei Chen, Li-Hua
  Sun, Jian-Jun Jia, Jin-Cai Wu, Xiao-Jun Jiang, Jian-Feng Wang, Yong-Mei
  Huang, Qiang Wang, Yi-Lin Zhou, Lei Deng, Tao Xi, Lu~Ma, Tai Hu, Qiang Zhang,
  Yu-Ao Chen, Nai-Le Liu, Xiang-Bin Wang, Zhen-Cai Zhu, Chao-Yang Lu, Rong Shu,
  Cheng-Zhi Peng, Jian-Yu Wang, and Jian-Wei Pan.
\newblock Satellite-to-ground quantum key distribution.
\newblock {\em Nature}, 549(7670):43--47, 2017.

\bibitem{Wengerowsky.2020}
S{\"o}ren Wengerowsky, Siddarth~Koduru Joshi, Fabian Steinlechner, Julien~R.
  Zichi, Bo~Liu, Thomas Scheidl, Sergiy~M. Dobrovolskiy, Ren{\'e} {van der
  Molen}, Johannes W.~N. Los, Val Zwiller, Marijn A.~M. Versteegh, Alberto
  Mura, Davide Calonico, Massimo Inguscio, Anton Zeilinger, Andr{\'e} Xuereb,
  and Rupert Ursin.
\newblock Passively stable distribution of polarisation entanglement over 192
  km of deployed optical fibre.
\newblock {\em npj Quantum Information}, 6(1):012307, 2020.

\bibitem{Marcikic.2006}
Ivan Marcikic, Ant{\'i}a Lamas-Linares, and Christian Kurtsiefer.
\newblock Free-space quantum key distribution with entangled photons.
\newblock {\em Applied Physics Letters}, 89(10):101122, 2006.

\bibitem{Shi.2020}
Yicheng Shi, Soe {Moe Thar}, Hou~Shun Poh, James~A. Grieve, Christian
  Kurtsiefer, and Alexander Ling.
\newblock Stable polarization entanglement based quantum key distribution over
  a deployed metropolitan fiber.
\newblock {\em Applied Physics Letters}, 117(12):124002, 2020.

\bibitem{Steinlechner.2017}
Fabian Steinlechner, Sebastian Ecker, Matthias Fink, Bo~Liu, Jessica Bavaresco,
  Marcus Huber, Thomas Scheidl, and Rupert Ursin.
\newblock Distribution of high-dimensional entanglement via an intra-city
  free-space link.
\newblock {\em Nature communications}, 8:15971, 2017.

\bibitem{Ursin.2007}
R.~Ursin, F.~Tiefenbacher, T.~Schmitt-Manderbach, H.~Weier, T.~Scheidl,
  M.~Lindenthal, B.~Blauensteiner, T.~Jennewein, J.~Perdigues, P.~Trojek,
  B.~{\"O}mer, M.~F{\"u}rst, M.~Meyenburg, J.~Rarity, Z.~Sodnik, C.~Barbieri,
  H.~Weinfurter, and A.~Zeilinger.
\newblock Entanglement-based quantum communication over 144 km.
\newblock {\em Nature Physics}, 3(7):481--486, 2007.

\bibitem{Ecker.2021}
Sebastian Ecker, Bo~Liu, Johannes Handsteiner, Matthias Fink, Dominik Rauch,
  Fabian Steinlechner, Thomas Scheidl, Anton Zeilinger, and Rupert Ursin.
\newblock Strategies for achieving high key rates in satellite-based qkd.
\newblock {\em npj Quantum Information}, 7(1), 2021.

\bibitem{Dierikx.2016}
Erik~F. Dierikx, Anders~E. Wallin, Thomas Fordell, Jani Myyry, Petri Koponen,
  Mikko Merimaa, Tjeerd~J. Pinkert, Jeroen C.~J. Koelemeij, Henk~Z. Peek, and
  Rob Smets.
\newblock White rabbit precision time protocol on long-distance fiber links.
\newblock {\em IEEE transactions on ultrasonics, ferroelectrics, and frequency
  control}, 63(7):945--952, 2016.

\bibitem{Wahl.2020}
Michael Wahl, Tino R{\"o}hlicke, Sebastian Kulisch, Sumeet Rohilla, Benedikt
  Kr{\"a}mer, and Andreas~C. Hocke.
\newblock Photon arrival time tagging with many channels, sub-nanosecond
  deadtime, very high throughput, and fiber optic remote synchronization.
\newblock {\em The Review of scientific instruments}, 91(1):013108, 2020.

\bibitem{Diamanti.2016}
Eleni Diamanti, Hoi-Kwong Lo, Bing Qi, and Zhiliang Yuan.
\newblock Practical challenges in quantum key distribution.
\newblock {\em npj Quantum Information}, 2(1):16025, 2016.

\bibitem{Oi.2017}
Daniel K.~L. Oi, Alex Ling, Giuseppe Vallone, Paolo Villoresi, Steve Greenland,
  Emma Kerr, Malcolm Macdonald, Harald Weinfurter, Hans Kuiper, Edoardo
  Charbon, and Rupert Ursin.
\newblock Cubesat quantum communications mission.
\newblock {\em EPJ Quantum Technology}, 4(1), 2017.

\bibitem{Kerstel.2018}
Erik Kerstel, Arnaud Gardelein, Mathieu Barthelemy, Matthias Fink,
  Siddarth~Koduru Joshi, Rupert Ursin, and {The CSUG Team}.
\newblock Nanobob: a cubesat mission concept for quantum communication
  experiments in an uplink configuration.
\newblock {\em EPJ Quantum Technology}, 5(1):6, 2018.

\bibitem{Ren.2017}
Ji-Gang Ren, Ping Xu, Hai-Lin Yong, Liang Zhang, Sheng-Kai Liao, Juan Yin,
  Wei-Yue Liu, Wen-Qi Cai, Meng Yang, Li~Li, Kui-Xing Yang, Xuan Han,
  Yong-Qiang Yao, Ji~Li, Hai-Yan Wu, Song Wan, Lei Liu, Ding-Quan Liu, Yao-Wu
  Kuang, Zhi-Ping He, Peng Shang, Cheng Guo, Ru-Hua Zheng, Kai Tian, Zhen-Cai
  Zhu, Nai-Le Liu, Chao-Yang Lu, Rong Shu, Yu-Ao Chen, Cheng-Zhi Peng, Jian-Yu
  Wang, and Jian-Wei Pan.
\newblock Ground-to-satellite quantum teleportation.
\newblock {\em Nature}, 549(7670):70--73, 2017.

\bibitem{Valivarthi.2016}
Raju Valivarthi, Marcel.li~Grimau Puigibert, Qiang Zhou, Gabriel~H. Aguilar,
  Varun~B. Verma, Francesco Marsili, Matthew~D. Shaw, Sae~Woo Nam, Daniel
  Oblak, and Wolfgang Tittel.
\newblock Quantum teleportation across a metropolitan fibre network.
\newblock {\em Nature Photonics}, 10(10):676--680, 2016.

\bibitem{Tsujimoto.2018}
Yoshiaki Tsujimoto, Motoki Tanaka, Nobuo Iwasaki, Rikizo Ikuta, Shigehito Miki,
  Taro Yamashita, Hirotaka Terai, Takashi Yamamoto, Masato Koashi, and Nobuyuki
  Imoto.
\newblock High-fidelity entanglement swapping and generation of three-qubit ghz
  state using asynchronous telecom photon pair sources.
\newblock {\em Scientific reports}, 8(1):1446, 2018.

\bibitem{Pan.1998}
Jian-Wei Pan, Dik Bouwmeester, Harald Weinfurter, and Anton Zeilinger.
\newblock Experimental entanglement swapping: Entangling photons that never
  interacted.
\newblock {\em Physical Review Letters}, 80(18):3891--3894, 1998.

\bibitem{Jozsa.2000}
Richard Jozsa, Daniel~S. Abrams, Jonathan~P. Dowling, and Colin~P. Williams.
\newblock Quantum clock synchronization based on shared prior entanglement.
\newblock {\em Physical Review Letters}, 85(9):2010--2013, 2000.

\bibitem{Chuang.2000}
Isaac~L. Chuang.
\newblock Quantum algorithm for distributed clock synchronization.
\newblock {\em Physical Review Letters}, 85(9):2006--2009, 2000.

\bibitem{Giovannetti.2001_disp}
V.~Giovannetti, S.~Lloyd, L.~Maccone, and F.~N. Wong.
\newblock Clock synchronization with dispersion cancellation.
\newblock {\em Physical review letters}, 87(11):117902, 2001.

\bibitem{Krco.2002}
Marko Kr{\v{c}}o and Prabasaj Paul.
\newblock Quantum clock synchronization: Multiparty protocol.
\newblock {\em Physical Review A}, 66(2), 2002.

\bibitem{Kong.2018}
Xiangyu Kong, Tao Xin, Shi-Jie Wei, Bixue Wang, Yunzhao Wang, Keren Li, and
  Gui-Lu Long.
\newblock Demonstration of multiparty quantum clock synchronization.
\newblock {\em Quantum Information Processing}, 17(11), 2018.

\bibitem{Giovannetti.2001}
Vittorio Giovannetti, Seth Lloyd, and Lorenzo Maccone.
\newblock Quantum-enhanced positioning and clock synchronization.
\newblock {\em Nature}, 412(6845):417--419, 2001.

\bibitem{Quan.2020}
Runai Quan, Ruifang Dong, Xiao Xiang, Baihong Li, Tao Liu, and Shougang Zhang.
\newblock High-precision nonlocal temporal correlation identification of
  entangled photon pairs for quantum clock synchronization.
\newblock {\em The Review of scientific instruments}, 91(12):123109, 2020.

\bibitem{Quan.2016}
Runai Quan, Yiwei Zhai, Mengmeng Wang, Feiyan Hou, Shaofeng Wang, Xiao Xiang,
  Tao Liu, Shougang Zhang, and Ruifang Dong.
\newblock Demonstration of quantum synchronization based on second-order
  quantum coherence of entangled photons.
\newblock {\em Scientific reports}, 6:30453, 2016.

\bibitem{Quan.2019}
Runai Quan, Ruifang Dong, Yiwei Zhai, Feiyan Hou, Xiao Xiang, Hui Zhou, Chaolin
  Lv, Zhen Wang, Lixing You, Tao Liu, and Shougang Zhang.
\newblock Simulation and realization of a second-order
  quantum-interference-based quantum clock synchronization at the femtosecond
  level.
\newblock {\em Optics letters}, 44(3):614--617, 2019.

\bibitem{DAuria.2020}
Virginia D'Auria, Bruno Fedrici, Lutfi~Arif Ngah, Florian Kaiser, Laurent
  Labont{\'e}, Olivier Alibart, and S{\'e}bastien Tanzilli.
\newblock A universal, plug-and-play synchronisation scheme for practical
  quantum networks.
\newblock {\em npj Quantum Information}, 6(1):1023, 2020.

\bibitem{Valencia.2004}
Alejandra Valencia, Giuliano Scarcelli, and Yanhua Shih.
\newblock Distant clock synchronization using entangled photon pairs.
\newblock {\em Applied Physics Letters}, 85(13):2655--2657, 2004.

\bibitem{Ho.2009}
Caleb Ho, Ant{\'i}a Lamas-Linares, and Christian Kurtsiefer.
\newblock Clock synchronization by remote detection of correlated photon pairs.
\newblock {\em New Journal of Physics}, 11(4):045011, 2009.

\bibitem{Yao.2012}
Yin-Ping Yao, Tong-Yi Zhang, Ren-Gang Wan, and Wei Zhao.
\newblock Review on quantum clock synchronization schemes.
\newblock In Pierre Galarneau, Xu~Liu, and Pengcheng Li, editors, {\em
  Photonics and Optoelectronics Meetings (POEM) 2011: Optoelectronic Sensing
  and Imaging}, SPIE Proceedings, page 833202. SPIE, 2012.

\bibitem{Lee.2019}
Jianwei Lee, Lijiong Shen, Alessandro Cer{\`e}, James Troupe, Antia
  Lamas-Linares, and Christian Kurtsiefer.
\newblock Symmetrical clock synchronization with time-correlated photon pairs.
\newblock {\em Applied Physics Letters}, 114(10):101102, 2019.

\bibitem{TiradoAndres.2019}
Francisco Tirado-Andr{\'e}s and Alvaro Araujo.
\newblock Performance of clock sources and their influence on time
  synchronization in wireless sensor networks.
\newblock {\em International Journal of Distributed Sensor Networks},
  15(9):155014771987937, 2019.

\bibitem{Bregni.1997}
S.~Bregni.
\newblock Clock stability characterization and measurement in
  telecommunications.
\newblock {\em IEEE Transactions on Instrumentation and Measurement},
  46(6):1284--1294, 1997.

\bibitem{CostantinoAgnesi.2020}
{Costantino Agnesi}, {Marco Avesani}, {Luca Calderaro}, {Andrea Stanco},
  {Giulio Foletto}, {Mujtaba Zahidy}, {Alessia Scriminich}, {Francesco
  Vedovato}, {Giuseppe Vallone}, and {Paolo Villoresi}.
\newblock Simple quantum key distribution with qubit-based synchronization and
  a self-compensating polarization encoder.
\newblock {\em Optica}, 7(4):284--290, 2020.

\bibitem{Calderaro.2020}
Luca Calderaro, Andrea Stanco, Costantino Agnesi, Marco Avesani, Daniele
  Dequal, Paolo Villoresi, and Giuseppe Vallone.
\newblock Fast and simple qubit-based synchronization for quantum key
  distribution.
\newblock {\em Physical Review Applied}, 13(5), 2020.

\bibitem{Williams.06.03.202112.03.2021}
James Williams, Martin Suchara, Tian Zhong, Hong Qiao, Rajkumar Kettimuthu, and
  Riku Fukumori.
\newblock Implementation of quantum key distribution and quantum clock
  synchronization via time bin encoding.
\newblock In Philip~R. Hemmer and Alan~L. Migdall, editors, {\em Quantum
  Computing, Communication, and Simulation}, page~5. SPIE, 06.03.2021 -
  12.03.2021.

\bibitem{GarciaLorenzo.2011}
B.~Garc{\'i}a-Lorenzo, A.~Eff-Darwich, J.~J. Fuensalida, and
  J.~Castro-Almaz{\'a}n.
\newblock Adaptive optics parameters connection to wind speed at the teide
  observatory: corrigendum.
\newblock {\em Monthly Notices of the Royal Astronomical Society},
  414(2):801--809, 2011.

\bibitem{Komar.2014}
P.~K{\'o}m{\'a}r, E.~M. Kessler, M.~Bishof, L.~Jiang, A.~S. S{\o}rensen, J.~Ye,
  and M.~D. Lukin.
\newblock A quantum network of clocks.
\newblock {\em Nature Physics}, 10(8):582--587, 2014.

\bibitem{Dai.2020}
Hui Dai, Qi~Shen, Chao-Ze Wang, Shuang-Lin Li, Wei-Yue Liu, Wen-Qi Cai,
  Sheng-Kai Liao, Ji-Gang Ren, Juan Yin, Yu-Ao Chen, Qiang Zhang, Feihu Xu,
  Cheng-Zhi Peng, and Jian-Wei Pan.
\newblock Towards satellite-based quantum-secure time transfer.
\newblock {\em Nature Physics}, 11:25, 2020.

\bibitem{Lee.2019_attack}
Jianwei Lee, Lijiong Shen, Alessandro Cer{\`e}, James Troupe, Antia
  Lamas-Linares, and Christian Kurtsiefer.
\newblock Asymmetric delay attack on an entanglement-based bidirectional clock
  synchronization protocol.
\newblock {\em Applied Physics Letters}, 115(14):141101, 2019.

\bibitem{Brunel.1999}
Christian Brunel, Brahim Lounis, Philippe Tamarat, and Michel Orrit.
\newblock Triggered source of single photons based on controlled single
  molecule fluorescence.
\newblock {\em Physical Review Letters}, 83(14):2722--2725, 1999.

\bibitem{Alleaume.2004}
R.~All{\'e}aume, F.~Treussart, J-M Courty, and J-F Roch.
\newblock Photon statistics characterization of a single-photon source.
\newblock {\em New Journal of Physics}, 6:85, 2004.

\bibitem{Martinez.2016}
L.~J. Mart{\'i}nez, T.~Pelini, V.~Waselowski, J.~R. Maze, B.~Gil, G.~Cassabois,
  and V.~Jacques.
\newblock Efficient single photon emission from a high-purity hexagonal boron
  nitride crystal.
\newblock {\em Physical Review B}, 94(12), 2016.

\bibitem{SamanthaIsaac.2020}
{Samantha Isaac}, {Andrew Conrad}, {Alex Hill}, {Kyle Herndon}, {Brian Wilens},
  {Dalton Chaffee}, {Daniel Sanchez-Rosales}, {Roderick Cochran}, {Daniel
  Gauthier}, and {Paul Kwiat}.
\newblock Drone-based quantum key distribution.
\newblock In {\em Conference on Lasers and Electro-Optics}, page JW2A.16.
  {Optical Society of America}, 2020.

\bibitem{Conrad.06.03.202112.03.2021}
Andrew Conrad, Samantha Isaac, Roderick Cochran, Daniel Sanchez-Rosales, Brian
  Wilens, Akash Gutha, Tahereh Rezaei, Dan Gauthier, and Paul Kwiat.
\newblock Drone-based quantum key distribution: Qkd.
\newblock In Hamid Hemmati and Don~M. Boroson, editors, {\em Free-Space Laser
  Communications XXXIII}, page~29. SPIE, 06.03.2021 - 12.03.2021.

\bibitem{Bruschi.2014}
David~Edward Bruschi, Timothy~C. Ralph, Ivette Fuentes, Thomas Jennewein, and
  Mohsen Razavi.
\newblock Spacetime effects on satellite-based quantum communications.
\newblock {\em Physical Review D}, 90(4), 2014.

\bibitem{Bruschi.2021}
David~Edward Bruschi, Symeon Chatzinotas, Frank~K. Wilhelm, and
  Andreas~Wolfgang Schell.
\newblock Spacetime effects on wavepackets of coherent light.
\newblock {\em Physical Review D}, 104(8), 2021.

\bibitem{Bruschi.20142}
David~Edward Bruschi, Animesh Datta, Rupert Ursin, Timothy~C. Ralph, and Ivette
  Fuentes.
\newblock Quantum estimation of the schwarzschild spacetime parameters of the
  earth.
\newblock {\em Physical Review D}, 90(12), 2014.

\bibitem{Kohlrus.2019}
Jan Kohlrus, David~Edward Bruschi, and Ivette Fuentes.
\newblock Quantum-metrology estimation of spacetime parameters of the earth
  outperforming classical precision.
\newblock {\em Physical Review A}, 99(3), 2019.

\bibitem{Ali.1998}
I.~Ali, N.~Al-Dhahir, and J.~E. Hershey.
\newblock Doppler characterization for leo satellites.
\newblock {\em IEEE Transactions on Communications}, 46(3):309--313, 1998.

\bibitem{Yin1140}
Juan Yin, Yuan Cao, Yu-Huai Li, Sheng-Kai Liao, Liang Zhang, Ji-Gang Ren,
  Wen-Qi Cai, Wei-Yue Liu, Bo~Li, Hui Dai, Guang-Bing Li, Qi-Ming Lu, Yun-Hong
  Gong, Yu~Xu, Shuang-Lin Li, Feng-Zhi Li, Ya-Yun Yin, Zi-Qing Jiang, Ming Li,
  Jian-Jun Jia, Ge~Ren, Dong He, Yi-Lin Zhou, Xiao-Xiang Zhang, Na~Wang, Xiang
  Chang, Zhen-Cai Zhu, Nai-Le Liu, Yu-Ao Chen, Chao-Yang Lu, Rong Shu,
  Cheng-Zhi Peng, Jian-Yu Wang, and Jian-Wei Pan.
\newblock Satellite-based entanglement distribution over 1200 kilometers.
\newblock {\em Science}, 356(6343):1140--1144, 2017.

\bibitem{Bergeron.2019}
Hugo Bergeron, Laura~C. Sinclair, William~C. Swann, Isaac Khader, Kevin~C.
  Cossel, Michael Cermak, Jean-Daniel Desch{\^e}nes, and Nathan~R. Newbury.
\newblock Femtosecond time synchronization of optical clocks off of a flying
  quadcopter.
\newblock {\em Nature communications}, 10(1):1819, 2019.

\bibitem{Troupe.27.01.201801.02.2018}
James Troupe and Antia Lamas-Linares.
\newblock Secure quantum clock synchronization.
\newblock In Zameer~U. Hasan, Philip~R. Hemmer, Alan~L. Migdall, and Alan~E.
  Craig, editors, {\em Advances in Photonics of Quantum Computing, Memory, and
  Communication XI}, page~20. SPIE, 27.01.2018 - 01.02.2018.

\bibitem{Zhao.2020}
Jie Zhao, Chaoxuan Ma, Michael R{\"u}sing, and Shayan Mookherjea.
\newblock High quality entangled photon pair generation in periodically poled
  thin-film lithium niobate waveguides.
\newblock {\em Physical review letters}, 124(16):163603, 2020.

\bibitem{He.2015}
Jiakun He, Bryn~A. Bell, Alvaro Casas-Bedoya, Yanbing Zhang, Alex~S. Clark,
  Chunle Xiong, and Benjamin~J. Eggleton.
\newblock Ultracompact quantum splitter of degenerate photon pairs.
\newblock {\em Optica}, 2(9):779, 2015.

\bibitem{C.Xiong.2011}
{C. Xiong}, {Christelle Monat}, {Alex S. Clark}, {Christian Grillet}, {Graham
  D. Marshall}, {M. J. Steel}, {Juntao Li}, {Liam O'Faolain}, {Thomas F.
  Krauss}, {John G. Rarity}, and {Benjamin J. Eggleton}.
\newblock Slow-light enhanced correlated photon pair generation in a silicon
  photonic crystal waveguide.
\newblock {\em Optics Letters}, 36(17):3413--3415, 2011.

\bibitem{Vanier.1981}
Jacques Vanier and Laurent-Guy Bernier.
\newblock On the signal-to-noise ratio and short-term stability of passive
  rubidium frequency standards.
\newblock {\em IEEE Transactions on Instrumentation and Measurement},
  IM-30(4):277--282, 1981.

\bibitem{Penrod.1996}
B.~M. Penrod.
\newblock Adaptive temperature compensation of gps disciplined quartz and
  rubidium oscillators.
\newblock In {\em Proceedings of 1996 IEEE International Frequency Control
  Symposium}, pages 980--987, 1996.

\end{thebibliography}
\bibliographystyle{unsrt}

\newpage
\appendix

\end{document}